\documentclass[aps,prd, twocolumn, amsmath,amssymb,showpacs,eqsecnum,nofootinbib]{revtex4}

\usepackage{graphicx}
\usepackage{dcolumn}
\usepackage{bm}

\begin{document}

\title{Hadamard Renormalization of the Stress Energy Tensor on the Horizons of a  Spherically Symmetric Black Hole Space-Time}
\author{Cormac Breen}
\email{cormac.breen@ucd.ie}
\author{Adrian C. Ottewill}
\email{adrian.ottewill@ucd.ie}
\affiliation{School of Mathematical Sciences and Complex \& Adaptive Systems Laboratory, University College Dublin, Belfield, Dublin 4, Dublin, Ireland}

\date{\today}

\begin{abstract}
We consider a quantum field which is in a Hartle-Hawking state propagating in a general spherically symmetric black hole space-time. We make use of uniform approximations to the radial equation to calculate the components of the stress tensor, renormalized using the Hadamard form of the Green's function, on the horizons of this  space-time. We then specialize these results to the case of the `lukewarm'  Reissner-Nordstrom-de Sitter black hole and derive some conditions on the stress tensor for the regularity of the Hartle-Hawking state.
\end{abstract}


\maketitle

\section{Introduction}
 The renormalized expectation value of the stress energy tensor operator $\langle \hat{T}_{\mu \nu}\rangle_{ren}$ is of fundamental importance in the study of quantum field theory in curved space time as it governs, via the semi-classical Einstein field equations
 \begin{align}
 G_{\mu \nu}+\Lambda g_{\mu \nu}= 8 \pi \langle \hat{T}_{\mu \nu}\rangle_{ren},
 \end{align}
 the back-reaction of the quantum field on the geometry of the space-time. There is a long and fruitful history of calculations of $\langle \hat{T}_{\mu \nu}\rangle_{ren}$ for various space-times, beginning with the work of Candelas and Howard for a massless, conformally coupled scalar field in the Schwarzschild black hole space-time \cite{Candelas,Howard:1984qp,Howard:1985yg}. This was then extended to the case of scalar fields with arbitrary mass and coupling to the Ricci scalar, in general spherically symmetric space-times, by Anderson, Hiscock and Samuel \cite{Anderson:1989vg, Anderson:1994hg}. Other examples of calculations of  $\langle \hat{T}_{\mu\nu} \rangle_{ren}$ can be found in \cite{DeBenedictis,Elster,Fawcett:1983dk,FrolovZel1,FrolovZel2,AndersonGroves,JensenOttewill:95}. 

 These calculations relied on the renormalization counterterms, denoted by  $\langle\hat{T}_{\mu \nu}\rangle_{DS}$, which were first calculated by Christensen \cite{Christensen:1976vb}. This calculation in turn relied on the DeWitt series representation for the Green's function, which is an asymptotic power series in inverse powers of the mass of the field $m$. This series is 
ill-defined for the massless case and requires some severe modification in order to be applicable to this case. We choose instead to follow the Hadamard renormalization procedure in the formulation of Brown and Ottewill \cite{BrownOttewill} which, building upon the axiomatic approach of Wald  \cite{Wald}, constructs a renormalization prescription which is well defined for both massive and massless fields. 

The calculation of the components of $ \langle \hat{T}_{\mu \nu}\rangle_{ren}$ in the exterior region of a black hole space-time naturally splits into two distinct parts, a quasi-analytical calculation to find the exact values on the horizons and a numerical calculation which is valid in the exterior region excluding the immediate vicinity of the horizon. This paper is concerned with the former calculation, while in an upcoming paper (henceforth referred to as Paper \textrm{II}), we will discuss how the application of the Hadamard renormalization procedure leads to an alternate method to that of Anderson, Hiscock and Samuel \cite{Anderson:1994hg}, for calculating $\langle \hat{T}_{\mu \nu}\rangle_{ren}$ in a general spherically symmetric space-time.

We note that these calculations do not assume that the exterior region of the space-time possess a single horizon, in fact, once the general calculations are done, we will apply the results to the case of a lukewarm black hole, whose exterior region possesses both an event and a cosmological horizon.

In a previous paper \cite{PhiSq}, we developed a technique for calculating the vacuum polarization of a scalar field on the horizon of a spherically symmetric black hole space-time. This paper was motivated by the numerical cancellation of horizon divergences to give a finite answer, which arose in a paper by Winstanley and Young on the vacuum polarization for lukewarm black holes \cite{Winstanley:2007}. It is therefore natural to ask if we may extend the technique developed in \cite{PhiSq} to the stress tensor case, allowing us to obtain finite values on the horizon without relying on numerical cancellations of divergent quantities.

This paper is organized as follows, in Sec.~\ref{sec:Had} we will outline the Hadamard renormalization procedure. In Sec.~\ref{sec:Hadap}  we will demonstrate the application of this method to obtain unrenormalized mode sum expressions and their renormalization subtraction counterterms. In Sec.~\ref{sec:Ren} we will calculate these mode sums using a uniform approximation to the radial solutions allowing us to obtain renormalized values which we then apply to the lukewarm case in Sec.~\ref{sec:LW}. Finally our conclusions are presented in Sec.~\ref{sec:Conclusions}. Throughout this paper we use the sign convention of Misner, Thorne and Wheeler \cite{MTW} and we will work in units in which $8\pi G=\hbar =c =k_B =1$.

\section{Hadamard Renormalization}
\label{sec:Had}
In this paper we will be concerned with the calculation of $\langle
\hat{T}_{\mu\nu}\rangle$ resulting from quantum scalar fields. The action for a scalar field $\phi$ with mass $m$ on a background space-time of dimension $n$, possessing metric $g_{\mu \nu}$, is given by
\begin{align}
S= -\frac{\sqrt{-g}}{2}\int d^{n} x \{ g^{\mu \nu}\phi_{,\mu} \phi_{,\nu}+ [m^2 +\xi R]\phi^2\},
\end{align} 
where $g$ denotes the determinant of $g_{\mu \nu}$, $R$ is the Ricci scalar and $\xi$ is a dimensionless constant describing the coupling between the scalar field and the gravitational field.
Requiring that the variation of the action with respect to $\phi$ vanishes yields the scalar field equation
\begin{align}
\label{Eq:field}
(\Box -m^2 -\xi R)\phi =0,
\end{align} 
where $\Box \equiv g^{\mu \nu} \nabla_{\mu} \nabla_{\nu}$.
The classical stress tensor is defined by the equation:
\begin{align}
\label{Eq:clas}
T^{\mu \nu} \equiv&2 g^{-1/2} \frac{\delta S}{\delta g_{\mu\nu}}\nonumber\\
=& (1-2 \xi) \phi^{;\mu}\phi^{;\nu} +(2 \xi -\tfrac{1}{2}) g^{\mu \nu} \phi_{;\alpha} \phi^{;\alpha} -2\xi  \phi  \phi^{;\mu \nu}  \nonumber\\&+2 \xi g^{\mu \nu} \phi \Box \phi 
+\xi(R^{\mu \nu}
-\tfrac{1}{2} R g^{\mu \nu})\phi^{2} -\frac{m^2}{2} g^{\mu \nu} \phi^2.
\end{align}
Vacuum expectation values of the various quadratic products of field operators can be identified with various Green's functions of the wave equation (\ref{Eq:field}). Of particular importance in the study of quantum field theory in curved space-time is the Feynman propagator $G_F$, defined as the time ordered product of the fields \cite{BD}
\begin{align}
&G_F(x,x')=i \langle 0|T( \hat{\phi}(x) \hat{\phi}(x')) | 0\rangle,\nonumber\\
=&i( \theta(t-t') \langle 0| \hat{\phi}(x) \hat{\phi}(x') | 0\rangle+ \theta(t-t') \langle 0| \hat{\phi}(x') \hat{\phi}(x) | 0\rangle).
\end{align}
The Feynman propagator satisfies the equation
\begin{align}
(\Box -m^2 -\xi R)G_F(x,x')= -\frac{\delta^{n}(x,x')}{\sqrt{g}}
\end{align}
where $\delta^{n}(x,x')$ is the $n$-dimensional delta function.

We will perform our calculations on a static spherically symmetric black hole background spacetime with line element:
\begin{equation}
\label{le}
\mathrm{d}s^2=-f(r) \mathrm{d} t^2 +\frac{1}{f(r)} \mathrm{d} r^2 +r^2 \mathrm{d}\theta^2 + r^2 \sin^2 \theta \mathrm{d}\phi^2 .
\end{equation}
The spacetime will have a horizon at $r=r_0$ whenever $f(r_0)=0$ and in this case the surface gravity of that horizon is given by $\kappa_0=|f'(r_0)|/2$ .
Following the standard procedure we Euclideanize our spacetime, that is we perform a Wick rotation $\tau \to i t$, Eq.~(\ref{le}) then becomes
\begin{equation}
\label{lee}
\mathrm{d}s^2=f(r) \mathrm{d} \tau^2 +\frac{1}{f(r)} \mathrm{d} r^2 +r^2 \mathrm{d}\theta^2 + r^2 \sin^2 \theta \mathrm{d}\phi^2.
\end{equation}
Assuming $\kappa_0 \neq 0$, this space will have a conical singularity whenever $f(r_0)=0$ which may be removed by making $\tau$ periodic with period $2\pi/\kappa_0$.
This periodicity in the Euclidean section corresponds in quantum field theory on the Lorentzian section to a thermal state with temperature
$T=\kappa_0/2\pi$.

In this configuration the quantum field $\hat{\phi}$ satisfies a wave equation where $\Box$ is now the Laplacian on Euclidean space. The Euclidean Green's function $G_E(x,x')$, for this elliptic equation, is related to $G_F$ by
\begin{align}
G_F(t,x;t',x')=-i G_E(i \tau,x ; i\tau',x'),
\end{align}
and satisfies the equation 
\begin{align}
\label{Eq:wavege}
(\Box -m^2 -\xi R)G_E(x,x')= -\frac{\delta^{4}(x,x')}{\sqrt{g}},
\end{align} 
with  $x=(\tau,r,\theta,\phi) $. A key advantage of Euclidean field theory is that $\Box$ is now an elliptic operator and hence has a unique, well defined inverse when supplemented by appropriate boundary conditions \cite{Garabedian}.\\
We may now define a formal expression for  $\langle\hat{T}_{\mu\nu}\rangle$ given by:
\begin{align}
\langle\hat{T}^{\mu\nu}\rangle=\mathcal{R}[ \lim_{x \to x'} \tau^{\mu \nu} G_E(x,x')],
\end{align}
where $\mathcal{R}$ denotes taking the real part, and $\tau^{\mu\nu}$ is a differential operator which reduces to Eq.~(\ref{Eq:clas}) in the coincidence limit, for example \cite{BrownOttewill}
\begin{align}
\label{Eq:tau}
\tau^{\mu \nu} =(1-2 \xi) g_{\nu'}^{~\nu} \nabla^{\mu}\nabla^{\nu'} +(2 \xi -\tfrac{1}{2}) g^{\mu\nu}g_{\alpha'}^{~\alpha}\nabla^{\alpha'}\nabla_{\alpha}\nonumber\\  - 2\xi \nabla^{\mu}\nabla^{\nu}  + 2 \xi g^{\mu\nu} \nabla_{\alpha}\nabla^{\alpha}+\xi(R^{\mu \nu} -\tfrac{1}{2} R g^{\mu\nu})-\tfrac{1}{2} m^2 g^{\mu\nu},  
\end{align}
where $g_{\nu'}^{~\nu}$ is the bivector of parallel transport, which serves to parallel transport a vector at $x'$ to a vector at $x$.

These expressions are perfectly valid in the classical theory, however in the quantum theory, being constructed from products of distributions evaluated at the same point, they are divergent. The characterization of these divergences (and their identification as an adjustment to the parameters of the theory) is the role of renormalization theory. As mentioned in the introduction the usual approach to renormalization begins with the DeWitt series representation. We choose to adopt an approach based upon the Hadamard series representation to the Euclidean Green's function, as it is equivalent to the standard method but is better defined, in particular, for the massless theory \cite{BrownOttewill}. We will now briefly outline this method.

In four dimensions the Green's function possesses the following singularity structure, first identified by  Hadamard  \cite{Hadamard,DeWittBrehme}: 
\begin{align}
\label{Eq:Had}
 G_E(x,x')= \frac{1}{8 \pi^2}\left[ \frac{\Delta^{1/2}}{\sigma} + V \ln(\lambda\sigma) +W\right], 
 \end{align}
where $2\sigma(x,x')$ denotes the square of the geodesic distance between $x$ and $x'$, and
\begin{align}
\Delta=-g^{-1/2}(x)\det(\sigma_{;\mu\nu'})g^{-1/2}(x')
\end{align}
is the biscalar form of the VanVleck-Morette determinant. $V(x,x')$ and $W(x,x')$ are regular biscalar functions possessing expansions of the form
\begin{align}
V(x,x')=\sum_{0}^{\infty} V_n(x,x')\sigma^{n};\quad W(x,x')=\sum_{0}^{\infty}W_n(x,x')\sigma^{n},
\end{align}
where $V_n(x,x')$ and $W_n(x,x')$ are themselves regular biscalar functions.  $\lambda$ is a constant which is required to ensure that the argument of the $\log$ function is dimensionless. Following the convention of Christensen \cite{Christensen:1976vb}, we let $\lambda= e^{2\gamma} \mu^2/2$, where $\mu=m$ for a massive field but is arbitrary when the field is massless \cite{Anderson:1994hg}. Following \cite{Anderson:1994hg}, we choose to set $\mu=1$ for massless fields.\\
Imposing Eq.~(\ref{Eq:wavege}) for $x\neq x'$, and demanding balance of explicit powers of $\sigma$. one can derive recursion relations for the coefficients $V_n(x,x')$ for $n \geq0$ and for $W_n(x,x')$ \cite{DeWittBrehme} for all $n >0$. $W_0(x,x')$ remains undetermined and corresponds to the freedom to add to $G_E(x,x')$ solutions of the homogeneous wave equation. The $V_n(x,x')$ are purely geometrical, all the information about the state in contained in $W(x,x')$. All the singular behaviour of the Green's function is geometrical and state independent.

We return now to our definition of the stress tensor
\begin{align}
\langle\hat{T}^{\mu\nu}\rangle=\lim_{x \to x'} \tau^{\mu\nu} G_E(x,x'),
 \end{align}
where $\tau^{\mu\nu}$ is defined in Eq.~(\ref{Eq:tau}) and it is implied that we are only taking the real part of the limit. As we have previously stated, this expression is divergent and is a priori meaningless. However,  as the singular behaviour is geometrical, the difference between the stress tensor of two different quantum states, $|A\rangle$ and $|B\rangle$ say, is well defined and given by
\begin{align}
&\langle A | \hat{T}^{\mu\nu}|A \rangle -\langle B | \hat{T}^{\mu\nu}|B \rangle \equiv \nonumber\\
&\lim_{x \to x'} \{ \tau^{\mu\nu} G_{EA}(x,x')- \tau^{\mu\nu} G_{EB}(x,x')\}\nonumber\\
=&[\tau^{\mu\nu}W_{A}(x,x')-\tau^{\mu\nu} W_{B}(x,x')],
 \end{align}
 where the square brackets denote that the coincidence limit has been taken. Now the tensor 
 \begin{align}
\tau^{\mu\nu}[ W_{A}]\equiv [\tau^{\mu\nu}W_{A}(x,x')]
 \end{align}
 is well defined and contains all the state dependent information. We will use it to construct a new definition for the stress tensor. Firstly we remark that although $\tau^{\mu\nu}[ W_{A}]$ is symmetric it is not,  in general, conserved. In fact it can be shown that \cite{BrownOttewill}
  \begin{align}
  \tau^{\mu\nu}{}_{;\nu}[ W_{A}]=-2v_1^{;\mu},
   \end{align}
   where $v_1\equiv [V_1(x,x')]$ and is given by the expression \cite{BrownOttewill}
\begin{align}
v_1 =\frac{1}{720}\left(R_{abcd}R^{abcd} -R_{ab}R^{ab}\right) -\frac{1}{24}(\xi -\tfrac{1}{5})\Box R\nonumber\\
+\frac{1}{8}(\xi -\tfrac{1}{6})^2 R^2 +\frac{1}{4} m^2(\xi -\tfrac{1}{6}) R+\frac{m^4}{8}.
\end{align}
If we therefore choose to define our renormalized stress tensor for a quantum state $|A\rangle$ by
 \begin{align}
 \label{Eq:Tnew}
\langle\hat{T}^{\mu\nu}\rangle_{ren}=\frac{1}{8\pi^2}\left(\tau^{\mu\nu}[ W_{A}]+ 2 v_1 g^{\mu\nu}\right),
 \end{align}
 then we see that our definition of the stress tensor is conserved and has trace \cite{BrownOttewill}
  \begin{align}
\langle\hat{T}^{~\mu}_{\mu}\rangle=\frac{1}{8\pi^2}\left(2v_1 +\tfrac{1}{2}(6\xi-1)\Box w_A(x) -m^2 w_A(x)\right).
 \end{align}
 where  $w_A\equiv [W_A(x,x')]$. In the conformal case this definition reproduces the standard trace anomaly $\displaystyle{v_1/4\pi^2}$.\\
 Wald has shown that if a given definition for a stress tensor satisfies certain natural conditions then it is unique up to the possible addition of conserved geometrical tensors corresponding to the metric variation of the terms in the gravational action  \cite{Wald}. The definition (\ref{Eq:Tnew}) satisfies these axioms \cite{BrownOttewill} and so is equivalent to any $\langle\hat{T}_{\mu\nu}\rangle_{ren}$ derived using one of the other renormalizing techniques, but has the added advantage of being more direct. As noted above we must introduce an implicit length scale in order to make the argument of the logarithm function in Eq.~(\ref{Eq:Had}) dimensionless. A different choice in length scale results in a different definition for $\langle\hat{T}_{\mu\nu}\rangle_{ren}$, however these tensors will differ only by a multiple of $ [\tau^{\mu\nu}V(x,x')]$, which is a conserved tensor in Wald's class.

Mc Laughlin has shown that renormalization using the Hadamard regularization procedure is equivalent to covariant geodesic point separation technique \cite{Mclaug}. In fact he has proven that, for scalar fields, the stress tensor obtained via Christensen's covariant geodesic point separation technique procedure is equal to that obtained using the Hadamard form plus a geometric term
\begin{align}
\label{Eq:mcalug}
\mathcal{M}^{\mu \nu}=\frac{m^2}{16 \pi^2} \left\{ \left(\xi -\frac{1}{6}\right)\left(R^{\mu \nu} -\frac{1}{2} g^{\mu \nu} R\right) -\frac{3}{8}m^2 g^{\mu \nu}\right\},
\end{align}
which can be absorbed into the left hand side of the semi classical Einstein equation in the usual way by renormalizing the constants $G$ and $\Lambda$. $\mathcal{M}^{\mu \nu}$ is clearly a conserved quantity, therefore we are free to add it to our definition of the stress tensor, without violating Wald's axioms, in order for our results to be in agreement with those obtained using Christensen's method. Hence we choose to give a final definition of the renormalized stress tensor as
\begin{align}
\langle\hat{T}^{\mu\nu}\rangle_{ren}=\frac{1}{8\pi^2}\left(\tau^{\mu\nu}[ W_{A}]+ 2 v_1 g^{\mu\nu}\right) + \mathcal{M}^{\mu \nu},
\end{align}
with trace
  \begin{align}
\langle\hat{T}^{~\mu}_{\mu}\rangle=\frac{1}{8\pi^2}\left(2v_1 +\tfrac{1}{2}(6\xi-1)\Box w_A(x) -m^2 w_A(x)\right) + \mathcal{M}^{~\mu}_{\mu},
 \end{align}
again giving the correct trace anomaly.\\
We note here that this analysis has been extended to arbitrary spacetime dimensions by D\'ecanini and Folacci~\cite{DecFol}.
\section{Application of Hadamard renormalization}
\label{sec:Hadap} 
In practice, we calculate $\tau^{\mu \nu}[ W_{A}]$ on a given space-time by subtracting off the action of $\tau^{\mu \nu}$ on the singular part of the Hadamard form (\ref{Eq:Had}) from an expression for the unrenormalized value of $\tau^{\mu \nu} G_E(x,x')$, then finally taking the coincidence limit. The unrenormalized expression we make use of was first derived for a Schwarzschild space-time by Candelas \cite{Candelas} and extended to the case of general spherically symmetric space-time by Anderson \cite{Anderson:1989vg} and is given by
\begin{align}
  \label{Eq:Ge1}
   G_{E}(x,x') =\frac{\kappa}{8 \pi^2} \sum_{n=-\infty}^{\infty}e^{i n \kappa (\tau -\tau')}\sum_{l=0}^{\infty} P_l(\cos\gamma) \chi_{nl}(r,r').
   \end{align}
 Where $n$ labels the mode's frequency, $l$ labels the mode's total angular momentum, and $\chi_{nl}(r,r')$ is the Green's function for the radial equation 
     \begin{align}
\label{Eq:1dwave}
 &\frac{1}{r^2}\frac{d}{dr} \left(r^2 f\frac{d \chi}{dr}\right)-\bigg(\frac{n^2 \kappa_0^2}{f} +\frac{l(l+1)}{r^2} +m^2 +\xi R\bigg) \chi\nonumber\\
& =-\frac{\delta(r-r')}{r^2}.
\end{align}
 We define $p_{nl}$ and $q_{nl}$ to be the independent solutions to the homogeneous version of this equation, 
with $p_{nl}$ defined to be the solution which is regular on the lower limit of the region under consideration, while $q_{nl}$ is regular at the upper limit of the region. $\chi_{nl}(r,r')$  is then given by 
\cite{Anderson:1989vg}
\begin{align}
\chi_{nl}(r,r')=C_{nl}p_{nl}(r_<)q_{nl}(r_>),
\end{align}
where $r_<$ is the lesser of the two points $r,r'$, and $r_>$ the greater. $C_{nl}$ is fixed by the Wronskian condition
\begin{align}
C_{nl}\left[p_{nl} \frac{dq_{nl}}{dr}-q_{nl} \frac{dp_{nl}}{dr}\right]=- \frac{1}{r^2 f}.
\end{align}
Since we have that $\chi_{nl}(r,r')$ is real, we may then express $G_{E}(x,x')$ for a thermal state in a form which we will use for the remainder of this paper:
\begin{align}
\label{Eq:modet}
G_E(x,x')=\sum_{n=0}^{\infty} F(n)\cos(n \kappa(\epsilon_{\tau}))\sum_{l=0}^{\infty}(2l+1)P_l(\cos\gamma)\nonumber\\
\times C_{nl} p_{nl}(r_<)q_{nl}(r_>),
\end{align}
where $\epsilon_{\tau}=\tau-\tau'$, $F(0)= \kappa/8 \pi^2$ and $F(n)=\kappa/4 \pi^2$, $n>0$. Henceforth we choose to set $r_>=r$. In static, spherically symmetric space-times, states respecting the same symmetries have a stress tenor $\langle \hat{T}^{\mu}_{~\nu}\rangle_{ren}$, which is diagonal. 
Using the definition given in the previous section we may express the diagonal elements of $\langle \hat{T}^{\mu}_{~\nu}\rangle_{ren}$ in the following manner:
\begin{align}
\label{Eq:Trenr}
\langle \hat{T}^{\nu}_{~\nu} \rangle_{ren}=2(\tfrac{1}{2} -\xi)[g^{\nu\nu'}G_{;\nu\nu'}]_{ren}+(2\xi -\tfrac{1}{2})[g^{\alpha\alpha'}G_{;\alpha \alpha '}]_{ren}
\nonumber\\
-2\xi[ g^{\nu\nu}G_{;\nu\nu}]_{ren}+2\xi [g^{\alpha \alpha}G_{;\alpha \alpha}]_{ren}+\xi(R^{\nu}_{~\nu} -\frac{1}{2} R)[G]_{ren}\nonumber\\
-\frac{m^2}{2}[G]_{ren}+\frac{2 v_1}{8 \pi^2} +\mathcal{M}^{\nu}_{~\nu}.
\end{align}
where $\nu$ is not summed over, ; denotes covariant differentiation and we have dropped the $(x,x')$ dependence as well as the $E$ subscript for notational convenience.

Our strategy for computing these components of $\langle \hat{T}^{\mu}_{~\nu}\rangle_{ren}$ is as follows. Firstly we calculate the required derivatives of $G_E(x,x')$, after which we may take the partial coincidence limit $\{t\to t', \theta \to \theta', \phi \to \phi'\}$ and also we place $r'$ on $r_0$. We then renormalize these expressions by subtracting off their respective singular Hadamard parts before taking the coincidence limit $r \to r_0$. Finally we insert these renormalized expressions into Eq.~(\ref{Eq:Trenr}).

\subsection{Unrenormalized mode sum expressions}
Before we can begin to consider working with Eq.~(\ref{Eq:Trenr}) we need to first calculate these bivectors for our particular choice of point separation.
The bivector of parallel transport are defined by the equation
\begin{equation}
\label{Eq:bivec}
\sigma^{; \alpha'}g_{a b' ; \alpha'}=0,
\end{equation}
with the boundary conditions that $g_{ab'} =g_{ab}$ when $x=x'$.
For a general point separation this expression can be quite complicated, however for radial separation this reduces to quite a simple form. By virtue of the symmetries of the space-time we have 
\begin{equation}
\sigma^{; \alpha'} = 0\quad \alpha \neq r',
\end{equation}
hence Eq.~(\ref{Eq:bivec}) reduces to
\begin{equation}
\sigma^{;r'}g_{a b' ;r'}=0.
\end{equation}
Using the definition of the covariant derivative we see that the bivectors of parallel transport for radial separation are determined by
\begin{equation}
\label{Eq:bivec2}
g_{a b' ,r'}=\Gamma^{\rho'}_{b' r'} g_{a \rho'}.
\end{equation}
 We require the non-vanishing components of the connection, these are
\begin{align}
&\Gamma^{r}_{r r}=-\frac{f'(r)}{2 f(r)} ; \quad \Gamma^{r}_{\theta \theta}= -r f(r); \quad \Gamma^{r}_{\phi \phi}= -r f(r) \sin^2\theta;\nonumber\\
&\Gamma^{r}_{tt}=\frac{f(r) f'(r)}{2}
\Gamma^{\theta}_{r \theta}=\Gamma^{\theta}_{\theta r}=\frac{1}{r};\quad
 \Gamma^{\theta}_{\phi \phi}=-\sin\theta \cos\theta;\nonumber\\
 & \Gamma^{\phi}_{r \phi}= \Gamma^{\phi}_{ \phi r}=\frac{1}{r}; \quad  \Gamma^{\phi}_{\theta \phi}= \Gamma^{\phi}_{ \phi \theta} =\cot \theta;
 \Gamma^{t}_{t r}=\Gamma^{t}_{ r t}=\frac{f'(r)}{2 f(r)}.
\end{align}
The calculation of the bivectors is  repetitive in nature so we will just show the calculation of the one and the rest follow similarly.\\
For $g_{rr'}$ we have:
\begin{align}
g_{rr' ,r'}=\Gamma^{\rho'}_{r' r'} g_{r \rho'}= \Gamma^{r'}_{r' r'} g_{r r'}=-\frac{f'(r')}{2f(r')}g_{rr'}.
\end{align}
Integration gives
\begin{equation}
g_{rr'}= \left(\frac{f(r)}{f(r')}\right)^{1/2} g_{rr} = \frac{1}{\sqrt{f(r)f(r')}}.
\end{equation}
The other components are given by:
\begin{equation}
g_{tt'}= \left(\frac{f(r')}{f(r)}\right)^{1/2} g_{tt} =-\sqrt{f(r)f(r')}.
\end{equation}
\begin{equation}
g_{\theta\theta'}= \frac{r'}{r}g_{\theta\theta}=rr'.
\end{equation}
\begin{equation}
g_{\phi\phi'}= \frac{r'}{r}g_{\phi\phi}=rr'\sin^2\theta.
\end{equation}
 It is straightforward to show that all other components of $g_{a b'}$ vanish.\\
We turn our attention to calculating the derivatives of $G_E(x,x')$. For the purposes of this calculation we choose to denote the partial coincidence limit, $\{t\to t', \theta \to \theta', \phi \to \phi'\}$, of a bitensor $A(x,x')$ by $\{A\}$. We will show the derivation details for one derivative $\{G_{;tt}\}$, the others follow along similar lines. 
Using the definition of the covariant derivative, we have that
 \begin{align}
G_{;tt} = G_{,tt} - \Gamma^{r}_{tt} G_{,r}.
 \end{align}
 We consider the partial derivative first
  \begin{align}
G_{,tt} &=-\frac{\partial^2}{ \partial \tau^2}\frac{\kappa}{4 \pi^2} \sum_{n=0}^{\infty} \cos(n\kappa\epsilon_\tau)\sum_{l=0}^{\infty}(2l+1)P_l(\cos\gamma)\nonumber\\
&\times p_{nl}(r')q_{nl}(r) \nonumber\\
& =\frac{\kappa}{4 \pi^2}\sum_{n=0}^{\infty} n^2 \kappa^2\cos(n\kappa\epsilon_\tau)\sum_{l=0}^{\infty}(2l+1)P_l(\cos\gamma)\nonumber\\
&\times p_{nl}(r')q_{nl}(r). 
 \end{align}
Where we have used the relation $\partial / \partial t=i \partial/ \partial\tau $. Taking the partial coincidence limit yields
  \begin{align}
\{g^{tt}G_{;tt}\}&=-
\frac{1}{f(r)}\frac{\kappa}{4 \pi^2}\sum_{n=1}^{\infty} n^2 \kappa^2\sum_{l=0}^{\infty}(2l+1)p_{nl}(r')q_{nl}(r)
 \end{align} 
This vanishes in the limit $r \to r_0$ as $p_{nl}(r_0) =0$ for $n >0$ as 
   \begin{align}
   \label{Eq:expanpnl}
   p_{nl}(r') &= a_0(r'-r_0)^{n/2}+O\left((r'-r_0)^{n/2+1}\right),
    \end{align} 
    with $a_0=1/\sqrt{\kappa r_0}$.
Likewise taking $r'\to r_0$  
  \begin{align}
\Gamma^{r}_{tt}\{G_{,r}\}= \frac{f(r)f'(r)}{2} \frac{a_0 \kappa}{4 \pi^2}\sum_{l=0}^{\infty}(2l+1)\frac{\textrm{d}q_{0l}}{\textrm{d}r},
 \end{align}
and hence
 \begin{align}
\{g^{tt}G_{;tt}\}&=\frac{f'(r)}{2} \frac{a_0 \kappa}{4 \pi^2}\sum_{l=0}^{\infty}(2l+1)\frac{\textrm{d}q_{0l}}{\textrm{d}r}.
 \end{align} 
Repeating this procedure for the other derivatives leads to the following mode sum expressions
\begin{align}
\label{Eq:moder}
\{G_E\}=\frac{\kappa}{8 \pi^2}a_0\sum_{l=0}^{\infty}(2l+1)q_{0l}(r).
\end{align}
  \begin{align}
\{g^{tt'}G_{;tt'}\}&=\frac{1}{\sqrt{f(r)}}\frac{\kappa^3}{4 \pi^2} \frac{a_0}{\sqrt{2 \kappa}}\sum_{l=0}^{\infty}(2l+1)q_{1l}(r) 
 \end{align} 
     \begin{align}
\{g^{rr'} G_{;r r'}\}
=\frac{\kappa}{4 \pi^2}\sqrt{\frac{f(r)\kappa}{2}} a_0 \sum_{l=0}^{\infty}(2l+1) \frac{\textrm{d}q_{1l}(r)}{\textrm{d}r} 
 \end{align} 
 \begin{align}
\frac{\{G_{;\theta \theta' }\}}{r^2}=  
\frac{\{G_{;\phi \phi'}\}}{r^2\sin^2 \theta}=\frac{\kappa a_0}{4 \pi^2}\sum_{l=0}^{\infty}(2l+1)\left(\frac{f(r)}{r}\frac{\textrm{d}q_{0l}(r)}{\textrm{d}r}\right.\nonumber\\
\left.-\frac{l(l+1)}{2r^2}q_{0l}(r)\right) \end{align}
  \begin{align}
\frac{\{G_{;\theta \theta' }\}}{r r_0}=  
\frac{\{G_{;\phi \phi'}\}}{r r_0 \sin^2 \theta}= \frac{\kappa a_0}{4 r_0 \pi^2}\sum_{l=0}^{\infty}(2l+1) \frac{l(l+1)}{2 r}q_{0l}(r)\
 \end{align} 
Note that we did not calculate a mode sum expression for $f(r)\{G_{;r r}\}$ since, as will be seen shortly, we may exploit the wave equation to obtain $[g^{rr}G_{r r}]_{ren}$ once we have calculated the renormaized value of the other components.
 
 \subsection{Renormalization subtraction terms}
 As  discussed earlier, the renormalization subtraction terms for a particular component are obtained by the action of the corresponding differential operator on the singular part of the Hadamard form of $G_{E}$, namely
\begin{equation}
\label{Eq:Gsing}
G_{sing} (x ,x') =\frac{1}{8 \pi^2} \left[ \frac{\Delta^{1/2}}{\sigma} + V \log(\lambda \sigma)\right].
\end{equation}
Using the expansion
\begin{align}
\label{Eq:coord}
\sigma=\frac{1}{2} g_{\alpha\beta}\Delta x^{\alpha}\Delta x^{\beta} + A_{\alpha\beta\gamma}\Delta x^{\alpha}\Delta x^{\beta}\Delta x^{\gamma} \nonumber\\
+  B_{\alpha\beta\gamma\delta}\Delta x^{\alpha}\Delta x^{\beta}\Delta x^{\gamma}\Delta x^{\delta}+\dots
\end{align}   
with
\begin{align}
A_{abc}=&-\frac{1}{4} g_{(ab,c)};\\
B_{abcd}=&-\frac{1}{3} \bigg( A_{(abc,d)} +
 g^{\alpha\beta}\bigg(\frac{1}{8} g_{(ab,|\alpha|}A_{|\beta|cd)}\nonumber\\ 
 &+\frac{9}{2}A_{\alpha(ab}A_{|\beta|cd)}\bigg)\bigg).
\end{align}
(the higher order terms are easily calculated by hand or by using Mathematica \cite{Mathematica}), 
Wardell and Ottewill  \cite{Barry:non-geo} have developed a Mathematica notebook which allows one to expand both $\Delta^{1/2}(x,x')$ and $V(x,x')$ in a coordinate expansion in powers of the coordinate separation of $\Delta x^{\alpha}$, for arbitrary point splitting to high order. 
 This notebook thus allows one to obtain a series expansion of $G_{sing} (x ,x')$ in powers of $\Delta x^{\alpha}$, up to the required order to capture both the divergence and the finite remainder terms of $G_{sing} (x ,x')$ and its derivatives in the coincidence limit.\\
 This method works perfectly for any regular point of the metric, indeed we will make use of this method in paper \textrm{II} to calculate the subtraction terms for temporal separation. In the case under consideration here, namely the case of radial separation about the horizon, we must modify this method slightly. This is due to the singularity, at $r=r_0$, of this coordinate system. This singularity is manifested through the function $f(r)$ which vanishes on the horizon, hence Eq.~(\ref{Eq:coord}), which is an expansion of $\sigma$ for small $\epsilon\equiv r-r_0$ and fixed $f(r)$, is no longer valid. To overcome this, we use the definition of $\sigma$ in terms of proper distance $s$:
\begin{equation}
\label{sig}
2\sigma = s^2,
\end{equation}
for space-like geodesics.
A great advantage of radial point splitting is that it allows one to integrate the line element to get an expression for $s$. Since $t=t',\theta=\theta'$ and $\phi=\phi'$, the line element becomes
\begin{equation*}
ds_{r}^2 =\frac{dr^2}{f(r)}
\end{equation*}
where $s_r$ denotes the proper distance along a radial geodesic. Hence
\begin{equation}
\label{int}
s_{r}=\int_{r'}^{r} \frac{1}{\sqrt{f(r')}} dr'.
\end{equation}
For a Ricci-flat space-time ($f(r)$ quadratic), one can perform this integral exactly and hence obtain an expression for $\sigma$ for radial splitting everywhere, without recourse to the expansion method. Unfortunately, for a general non Ricci-flat space-time, this is not the case. We can, however, expand the integrand about the horizon, then integration yields an expression for $s_r$ in terms of $\epsilon\equiv r-r_0$. Using this relation we may obtain an expression for $\sigma$ and hence $\Delta^{1/2}(x,x')$ and $V(x,x')$ which are now valid for radial separation about the horizon. Implementation of the method of Wardell and Ottewill, adapted in the manner just outlined, leads to the expressions listed below for the required subtraction terms. 
We note here that for simplicity, we choose to perform our calculations for a space-time with a constant Ricci scalar $R$. Therfore we may define an effective field mass $\hat{m}=\sqrt{m^2 +(\xi -1/6)R}$. It is worth noting also that space-times with a cosmological constant that are solutions of Einstein's vacuum equations, possess a constant Ricci scalar. 
 \begin{widetext}
\begin{align}
\{G_{sing}\}=\frac{f'_0}{16 \pi^2 \epsilon} 
+\frac{\hat{m}^2}{16 \pi^2}\ln\bigg(\frac{\lambda \epsilon}{f'_0}\bigg) 
 -\frac{f'_0}{48\pi^2 r_0}+O(\epsilon \ln(\epsilon))
\end{align}
\begin{align}
          \label{Eq:Gttdiv}
\{g^{tt}G_{Esing;tt}\}=-\frac{\kappa^2}{8\pi^2 \epsilon^2}
+ \frac{\kappa}{16 \pi^2 \epsilon} \left(\hat{m}^2- f_0''\right)+F_{tt}
+\frac{1}{11520 \pi^2 r_0^4}\left\{-4 f_0''{}^2r_0^2+2 f'_0
   r_0^3 \left(2 f''_0+f'''_0
   r_0-60 \hat{m}^2\right)\right.\nonumber\\
   \left.-r_0^4
   \left(f''_0{}^2+60 f''_0 \hat{m}^2-180
   \hat{m}^4\right)+4\right\}\ln\left( \epsilon\right) +O(\epsilon \ln(\epsilon)),
                \end{align}
        \begin{align}
        \label{Eq:Gththdiv}
      &  \{g^{\theta\theta}G_{sing;\theta\theta}\}=\{ g^{\phi\phi}G_{sing;;\phi\phi}\}=-\frac{\kappa^2}{8\pi^2 \epsilon^2}
 + \frac{\kappa}{96 \pi^2 \epsilon} \left(R +6\hat{m}^2\right)+F_{\theta\theta}  \nonumber\\
&+\frac{1}{11520 \pi^2 r_0^4}\left\{{4 f^{'2}_0r_0^2-2 f'_0 r_0^3 \left(2
   f''_0+f'''_0 r_0+60
   \hat{m}^2\right)+r_0^4 \left(f^{''2}_0+180
   \hat{m}^4\right)+120 \hat{m}^2r_0^2-4}\right\}\ln(\epsilon)+O(\epsilon \ln(\epsilon))
\end{align}
\begin{align}
 \label{Eq:Grrpdiv}
&\{g^{rr'}G_{sing;rr'}\}= -\frac{3\kappa^2}{8\pi^2 \epsilon^2}  + \frac{\kappa}{16 \pi^2 \epsilon} \left(\hat{m}^2-f''_0\right)+F_{rr'}\nonumber\\
&+\frac{1}{11520 \pi ^2 r_0^4}\left\{r_0^4 f''_0 \left(f''_0+60 \hat{m}^2\right)+4
   r_0^2 f^{'2}_0-2 r_0^3 f'_0\left(r_0 f'''_0)
2 f''_0-60 \hat{m}^2\right)-4
   \left(45 \hat{m}^4 r_0^4+1\right)\right\}\ln\left(\epsilon\right)+O(\epsilon \ln(\epsilon))
  \end{align}
    \begin{align}
  \label{Eq:Gttpdiv}
&\{g^{tt'}G_{sing;tt'}\}= \frac{\kappa^2}{8\pi^2 \epsilon^2}  - \frac{\kappa \hat{m}^2}{16 \pi^2 \epsilon} +F_{tt'}\nonumber\\
&+\frac{1}{11520 \pi ^2 r_0^4}\left\{r_0^4 f''_0 \left(f''_0+60 \hat{m}^2\right)+4
   r_0^2 f^{'2}_0\right.
  \left.-2 r_0^3 f'_0\left(r_0 f'''_0)
2 f''_0-60 \hat{m}^2\right)-4
   \left(45 \hat{m}^4 r_0^4+1\right)\right\}\ln\left(\epsilon\right)+O(\epsilon \ln(\epsilon))
  \end{align}
          \begin{align}
          \label{Eq:Gththpdiv}
&\{g^{\theta\theta'}G_{sing;\theta\theta'}\}=\{g^{\phi\phi'}G_{sing;\phi\phi'}\}=\frac{\kappa^2}{8\pi^2 \epsilon^2} - \frac{\kappa}{16 \pi^2 \epsilon} \left(\hat{m}^2 -\tfrac{1}{6} f''_0 +\tfrac{1}{3 r_0^2}\left(1+r_0 f'_0\right)\right)+F_{\theta \theta'}  \nonumber\\
&+\frac{1}{11520 \pi ^2 r_0^4}\left\{-4 r_0^2 \left(f^{'2}_0+30 \hat{m}^2\right)-r_0^4
   \left(-2 f'_0f'''_0+f^{'2}_0+180
   \hat{m}^4\right)+4f'_0 r_0^3 \left(f''_0+30
   \hat{m}^2\right)+4\right\}\ln\left(\epsilon\right) +O(\epsilon \ln(\epsilon)),
\end{align}
 where
     \begin{align}
     \label{Eq:Ftt}
     F_{tt}=&\frac{1}{11520 \pi ^2 r_0^4}\left\{\ln \left(\frac{\lambda}{f'_0}\right) \left[-4 f_0''{}^2r_0^2+2 f'_0 r_0^3 \left(2 f''_0+f'''_0
   r_0-60 \hat{m}^2\right)-r_0^4 \left(\left(f^{''2}_0+60 f''_0 \hat{m}^2-180 \hat{m}^4\right)\right)+4\right]\right.\nonumber\\
  & \left.-4
   f_0''{}^2r_0^2 (\ln (2)-15)+f'_0 r_0^3 \left(4 f''_0 (\ln (2)-3)+2 f'''_0 r_0 (\ln
   (2)-92)-120 \hat{m}^2 (1+\ln (2))\right)\right.\nonumber\\
 &  \left.+r_0^4 \left(-f^{''2}_0 (\ln (2)-2)-60 f''_0\hat{m}^2 (\ln (2)-4)+180 \hat{m}^4
   (1+\ln (2))\right)+4+\ln (16)\right\}.
        \end{align}
        \begin{align} 
  \label{Eq:Fthth}
 F_{\theta\theta}=\frac{1}{11520 \pi ^2 r_0^4}\left\{\ln \left(\frac{\lambda}{f'_0}\right)\left[4 r_0^2 \left(f^{'2}_0+30 \hat{m}^2\right)+r_0^4
   \left(-2 f'_0 f'''_0 +f^{''2}_0+180 \hat{m}^4\right)-4
 f'_0 r_0^3 \left( f''_0+30 \hat{m}^2\right)-4\right]\right.\nonumber\\
\left.+r_0 \left[4 f_0''{}^2r_0 (39+\ln (2))-f'_0
   \left(r_0^2 \left(f''_0 (248+\ln (16)+f'''_0
   r_0 (4+\ln (4))+120 \hat{m}^2 (\ln
   (2)-1)\right)+280\right)\right.\right.\nonumber\\
   \left.\left.+f^{''2}_0 r_0^3 (2+\ln (2))
+60
   \hat{m}^2 r_0 \left(\hat{m}^2 r_0^2 (3+\ln (8))+\ln
   (4)\right)\right]+4(1-\ln (2))\right\}
 \end{align}
 \begin{align}
   \label{Eq:Frrp}
 F_{rr'}=\frac{1}{11520 \pi ^2 r_0^4}\left\{\ln \left(\frac{\lambda}{f'_0}\right)\left[ r_0^4 f''_0 \left(f''_0+60 \hat{m}^2\right)+4
   r_0^2 f^{'2}_0-2 r_0^3 f'_0
   \left(r_0 f'''_0+2 f''_0-60 \hat{m}^2\right)-4
   \left(45 \hat{m}^4 r_0^4+1\right)\right]\right.\nonumber\\
   \left.+4 f_0''{}^2r_0^2 (2+\ln (2))+f^{''2}_0 r_0^4 \ln
   (2)+60 f''_0 \hat{m}^2 r_0^4 (4+\ln (2))-4 (3+\ln (2))
   \left(45 \hat{m}^4 r_0^4+1\right)\right.\nonumber\\
   \left.-2 f'_0 r_0^3
   \left[f''_0 (28+\ln (4)+f'''_0r_0 (45+\ln
   (2))-60 \hat{m}^2 (3+\ln (2))\right]\right\}
  \end{align}
    \begin{align}
   \label{Eq:Fttp}
 F_{tt'}=\frac{1}{11520 \pi ^2 r_0^4}\left\{\ln \left(\frac{\lambda}{f'_0}\right)\left[ r_0^4 f''_0 \left(f''_0+60 \hat{m}^2\right)+4
   r_0^2 f^{'2}_0-2 r_0^3 f'_0
   \left(r_0 f'''_0+2 f''_0-60 \hat{m}^2\right)-4
   \left(45 \hat{m}^4 r_0^4+1\right)\right]\right.\nonumber\\
   \left.+f_0''{}^2r_0^2 \ln (16)+f^{''2}_0 r_0^4 (\ln
   (2)-2)+60 f''_0 \hat{m}^2 r_0^4 (2+\ln (2))-4 (1+\ln (2))
   \left(45 \hat{m}^4 r_0^4+1\right)\right.\nonumber\\
   \left.-2 f'_0r_0^3
   \left[f''_0 (24+\ln (4))+f''_0 r_0 (13+\ln
   (2))-60 \hat{m}^2 (1+\ln (2))\right]\right\}
  \end{align}
\begin{align}
  \label{Eq:Fththp}
 F_{\theta\theta'}=\frac{1}{11520 \pi ^2 r_0^4}\left\{\ln \left(\frac{\lambda}{f'_0}\right)\left[ -4 r_0^2 \left(f^{'2}_0+30 \hat{m}^2\right)-r_0^4
   \left(-2 f'_0f'''_0+f^{'2}_0+180
   \hat{m}^4\right)+4f'_0 r_0^3 \left(f''_0+30
   \hat{m}^2\right) +4\right] \right.\nonumber\\
   \left.+r_0 \left[-4 f_0''{}^2r_0 (\ln
   (2)-21)+f'_0\left\{r_0^2 \left[f''_0 (\ln
   (16)-52)+f'''_0 r_0 (4+\ln (4))+120 \hat{m}^2 (2+\ln
   (2))\right]+160\right\}\right.\right.\nonumber\\
\left.\left.-r_0 \left(f^{''2}_0 r_0^2
   (2+\ln (2))+60 \hat{m}^2 \left(\hat{m}^2 r_0^2 (3+\ln (8))+\ln
   (4)\right)\right)\right]-4+\ln (16)\right\}
\end{align}
 \end{widetext}
The extension of the above calculation to non-Ricci constant space times is straightforward.
\section{Renormalized Horizon Values}
\label{sec:Ren}
In this section we introduce uniform approximations to the radial function $q_{nl}$, which will allow us to calculate the unrenormalized mode sums to the required order. We then expand the resulting
expression in the near horizon limit and show that the divergent terms will cancel with those contained in the counterterms calculated in the previous section. Finally we will the take the horizon limit leaving us with the renormalized horizon values we desired.
\subsection{Approximations of $q_{0l}(r)$ and $q_{1l}(r)$}
In a previous paper \cite{PhiSq}, we demonstrated that the EGL uniform approximation to $q_{0l}(r)$ captured enough of the horizon behaviour to facilitate the calculation of the vacuum polarization on the horizons of a spherically symmetric black hole space-time. In order to perform the corresponding calculations for the stress tensor, we require an uniform approximation which captures more of the horizon behaviour. The analysis contained in \cite{PhiSq} can be extended, giving uniform approximations to both $p_{nl}(r)$ and $q_{nl}(r)$  in terms of Whittaker functions. For the purposes of this paper, we require only the zeroth order approximation to $q_{nl}(r)$, which is given by:
\begin{align}
Q^{W(0)}_{nl}(r)=&\frac{F_2(n)}{(r^2 f(r) \xi)^{1/4}}W_{-\nu_n,n/2}(2\sqrt{\psi_n} \xi)  \nonumber\\
F_2(n)&=\begin{cases}\displaystyle{ \frac{\Gamma(\tfrac{1}{2} +\nu_n)}{2^{3/2}\psi_n^{1/4}}}& n=0\\
 \displaystyle{\frac{\Gamma(\tfrac{1}{2} +n/2+\nu_n)}{(2\kappa r_0^2)^{n/2}2^{1/2}\psi_n^{(1-n)/4}}} & n>0.
\end{cases}
\end{align}
Here $\Gamma$ is the gamma function, $W$ is a Whittaker function of the second kind and 
\begin{align}
\label{Eq:Q0def}
 & \xi= \left(\int_{r_0}^{r} \frac{dr'}{\sqrt{r'^2f(r')}}\right)^2; \qquad  \nu_n=\frac{k_n^2}{8\sqrt{\psi_n}};\nonumber\\
 & k_n^2 =V_0 -\frac{R_0 r_0^2}{6} +\frac{1}{3}(n^2-1) +n^2\left(\frac{R_0 r_0^2}{6} + 2 \kappa r_0\right)\nonumber\\
& \psi_n =\frac{1}{960} r_0 \left[f'_0 \left(-3 \left(n^2-4\right) r_0^3 f'''_0\right.\right.\nonumber\\
&\left.\left.+8 r_0
   f'_0 \left(2 n^2-60 \xi +7\right)+120 \left(m^2 r_0^2+2 \xi \right)\right)\right.\nonumber\\
 &  \left.-8 r_0^2
   f''_0 f'_0 \left(2 n^2+15 \xi -8\right)+\left(4 n^2-1\right) r_0^3
   f^{''2}_0\right]
 \end{align}
with $V_0 =l(l+1) +(m^2 +\xi R_0)r_0^2$.\\
We will now demonstrate that the zeroth order approximation captures enough of the near horizon behaviour of $q_{0l}(r)$ to facilitate the caculation of the required mode sums. We will show the details for the $n=0$ approximation.\\ 
We begin by noting that 
using standard Frobenius theory we can find a series expansion for $q_{0l}$ about the event horizon.
To do this, we require the following expansions in $\epsilon$ (with $V_{0l}(r) = l(l+1) +(m^2 +\xi R)r^2$):
\begin{subequations}
\begin{equation}
\label{Eq:expanp}
\epsilon\frac{(r^2f(r))'}{r^2 f(r)} = p_0 +p_1\epsilon +p_2 \epsilon^2 +O(\epsilon^3) ;
\end{equation}
\begin{equation}
\label{Eq:Vexpan}
\epsilon^2\frac{V_{0l}(r)}{r^2 f(r)} = q_1\epsilon +q_2 \epsilon^2 +O(\epsilon^3);
\end{equation}
\end{subequations}
with
\begin{align}
\label{Eq:qpforbcof}
&p_0=1; \qquad p_1=\frac{f''_0}{4 \kappa} +\frac{2}{r_0};  \qquad p_2= -\frac{f_0''^2}{16\kappa^2}-\frac{2}{r_0^2}+\frac{f'''_0}{6 \kappa};\nonumber\\
&q_1= -\frac{V_0}{2\kappa r_0^2}; \qquad q_2=\frac{l(l+1)}{\kappa r_0^3} +\frac{f''_0}{8 \kappa^2 r_0^2} V_0,
\end{align}
The irregular solution obtained by Frobenius analysis on the radial equation near the regular singular point $r=r_0$ then has the form ~\cite{Burkill} 
\begin{align}
\label{Eq:qseries}
&q_{0l}(r) = -\frac{a_0}{2}\left[ \left(1 -q_1 \epsilon +\frac{q_1(p_1 +q_1) -q_2}{4} \epsilon^2\right) \ln(\epsilon) \right.\nonumber\\
&\left.+\left((2q_1 -p_1)\epsilon+\frac{p_1^2 -p_2 -p_1q_1 -3q_1^2 +q_2}{4} \epsilon^2\right)\right] \nonumber\\
&+O(\epsilon^3\ln(\epsilon)),
\end{align}
We may form the series for the solution $q_{0l}(r)$ near $r=r_0$ by adding an appropriate multiple of the regular solution, $\alpha_l p_{0l}(r)$, to ensure its satisfies the boundary conditions at the outer boundary, for example, regularity on the cosmological horizon for a lukewarm black hole or vanishing on the outer boundary if the black hole is contained in a reflecting box \cite{DuffyOt}. We note here that since our approximation, $Q^{W(0)}_{0l}$, is a local approximation, it cannot contain all the global information contained in the full solution, i.e it will differ form the full solution in its $\alpha_{0l}$ term.\\
Next, we consider the zeroth order approximation to $q_{0l}$, $Q^{W(0)}_{0l}$
\begin{equation}
\label{Eq:Whit}
Q^{W(0)}_{0l}(r) =\frac{1}{(\xi r^2 f)^{1/4}}\frac{\Gamma\left(\textstyle{\frac{1}{2} -\nu}\right)}{2^{3/2}( \psi_ 0)^{1/4}}W_{-\nu_0,0} \left(2\sqrt{\psi_0} \xi\right)
\end{equation}
It is straightforward to show that (\ref{Eq:Whit}) satisfies
\begin{align}
\label{Eq:W}
\frac{d}{dr}\left(r^2 f(r)\frac{d}{dr}Q^{W(0)}_{0l}(r)\right) -\tilde{V}^{W}_{0l}(r)Q^{W(0)}_{0l}(r)=0,
\end{align}
with
\begin{align*}
\tilde{V}^{W}_{0l}(r) =k_0^{2} -\frac{1}{4 \xi(r)} -\frac{f}{4} +\frac{r^2 f'^2}{16 f} -\frac{r(3 f' +r f'')}{4}\nonumber\\
 + 4 \psi_0 \xi(r).
\end{align*}
We now wish to apply Frobenius theory to Eq.~(\ref{Eq:W}). To do this analysis we need the equivalent of  the expansions (\ref{Eq:expanp}) and (\ref{Eq:Vexpan}). Clearly the expansion (\ref{Eq:expanp}) is the same in this case, so all we need is to find the equivalent of  Eq.~(\ref{Eq:Vexpan}) for $\tilde{V}(r)$. This is readily computed and we find that
\begin{equation}
\label{Eq:Vtilexpan}
\epsilon^2\frac{\tilde{V}^{W}_{0l}(r)}{r^2 f(r)} = q_1\epsilon +q_2 \epsilon^2 +O(\epsilon^3),
\end{equation}
with $q_1$ and $q_2$ given by Eq.~(\ref{Eq:qpforbcof}) i.e. the potentials $V_{0l}$ and $\tilde{V}^{W}_{0l}$ agree to this order.
Hence, we can conclude that the local terms of the series expansion of $Q_{0}(r)$ about the horizon is in agreement with Eq.~(\ref{Eq:qseries}) up to $O\left(\epsilon^3\ln(\epsilon)\right)$ and, as discussed above, the global terms $\alpha_l$ term will differ. So we may then write
\begin{equation}
\label{Eq:asym}
q_{0l}(r)= Q^{W(0)}_{0l}(r) +\beta^{W}_{0l} p_{0l}(r) +\mathcal{R}^{W}_{0l}(r),
\end{equation}
where $\mathcal{R}^{W}_{0l}(r)$ denotes the remainder terms and is $O(\epsilon^3\ln(\epsilon))$ as $\epsilon \to 0$. The constants $\beta_{0l}$ are determined by the requirement that $q_{0l}(r)$ is regular on the outer boundary  (i.e. contains the correct multiple of the regular solution) and are given in Appendix~\ref{Ap:Beta}.\\
We wish to use this approximation to calculate the summations
 \begin{align}
 S_1(r)=  \sum_{l=0}^{\infty}(2l+1)\frac{l(l+1)}{2}q_{0l}(r),\nonumber\\
  S_2(r)= f'(r)\sum_{l=0}^{\infty}(2l+1)\frac{dq_{0l}(r)}{dr}.
 \end{align}
It can shown that for $l\geq\epsilon^{-1/2}$, the combination $\beta^{W}_{0l} p_{0l}(r) +\mathcal{R}^{W}_{0l}(r)$ cuts off exponentially in $l$. Therefore we have an asymptotic expression for $S_1(r)$  and $S_2(r)$ valid in the region of the horizon:
\begin{align}
S_1(r)=  \sum_{l=0}^{\infty}(2l+1)\frac{l(l+1)}{2}Q^{W(0)}_{0l}(r) +O(\epsilon \ln(\epsilon)),\nonumber\\
S_2(r)= f'(r)\sum_{l=0}^{\infty}(2l+1)\frac{dQ^{W(0)}_{0l}(r)}{dr}+O(\epsilon \ln(\epsilon)),
\end{align} 
and so we may conclude that the approximation $Q^{W(0)}_{0l}(r)$ is of sufficient accuracy for the calculation of the $n=0$ contribution to the stress tensor.
Similar analysis for the $n=1$ approximation $Q^{W(0)}_{1l}(r)$ allows us to conclude that
\begin{equation}
\label{Eq:asym1}
q_{1l}(r)= Q^{W}_{1l}(r) +\beta^{W}_{1l} p_{1l}(r) +R^{W}_{1l}(r),
\end{equation}
where $R_{1l}(r)$ denotes the remainder terms which are $O\left(\epsilon^{5/2}\right)$ as $\epsilon \to 0$. The constants $\beta^W_{1l}$ are determined by the requirement that $q_{1l}(r)$ is regular on the outer boundary and are given in Appendix~\ref{Ap:Beta}.
For the $n=1$ mode we are required to calculate the following sums for the stress tensor:
  \begin{align}
S_{3}=\frac{1}{\sqrt{f(r)}} \sum_{l=0}^{\infty}(2l+1)q_{1l}(r);\nonumber\\
 S_{4} =\sqrt{f} \sum_{l=0}^{\infty}(2l+1)\frac{d q_{1l}(r)}{dr}.
\end{align}
 The argument proceeds along the same lines as in the  case of $q_{0l}$, allowing us to conclude that 
  \begin{align}
S_{3}=\frac{1}{\sqrt{f(r)}} \sum_{l=0}^{\infty}(2l+1)Q^{W}_{1l}(r)+O(\epsilon),
\end{align} 
\begin{align}
S_{4} =\sqrt{f} \sum_{l=0}^{\infty}(2l+1)\frac{d Q^{W}_{1l}(r)}{dr} +O(\epsilon),
\end{align}
The calculation of the contributing sums over $l$ of $\beta^{W}_{0l}$ and $\beta^{W}_{1l}$ present no problem as they can be shown numerically to converge like $l^{-3}$.

\subsection{Mode Sum Calculation}
These calculations are tedious and repetitive in nature, so for the sake of brevity we will outline the calculation details of one component, $[g^{\theta\theta'}G_{\theta \theta'}]_{ren}$,  and simply list the results for the other components. The interested reader may refer to \cite{PHD} for a more detailed discussion.
We wish to compute the renormalized value,  on the black hole horizon $r=r_0$, of 
\begin{align}
\label{Eq:Gththp}
\{ g^{\theta\theta'}G_{;\theta \theta'}\}= \frac{\sqrt{\kappa}}{8\pi^2 r_0^2} \frac{S_1}{r}
 \end{align}
where again,
\begin{align}
S_1(r)=  \sum_{l=0}^{\infty}(2l+1)\frac{l(l+1)}{2}Q^{W(0)}_{0l}(r) +O(\epsilon \ln(\epsilon)).
\end{align}
Unfortunately, we have no way of computing this sum in its current form, however we may rexpress it as
\begin{align}
S_1&=  \sum_{l=0}^{\infty}(2l+1)\frac{l(l+1)}{2} \nonumber\\
&\times\left[Q^{W(0)}_{0l}(r)-\left(\frac{\xi}{r^2 f}\right)^{1/4}K_0(k_0 \xi^{1/2}) \right]\nonumber\\
&+\sum_{l=0}^{\infty}(2l+1)\frac{l(l+1)}{2}\left(\frac{\xi}{r^2 f}\right)^{1/4}K_0(k_0 \xi^{1/2}) \nonumber\\
&=S_{1a} +S_{1b},
\end{align}
where, as above $ k_0^2 =V_0 -\frac{1}{6}R_0 r_0^2 +\frac{1}{3}$. We will now proceed to calculate the two component sums in the above expression, beginning with $S_{1b}$ as it is the most straightforward.

\paragraph{\bf Evaluation of $S_{1b}$\\}
We may calculate $S_{1b}$ by making use of the Watson-Sommerfeld formula, giving
\begin{align}
S_{1b} &= \int^{\infty}_{0} \lambda(\lambda^2 -\tfrac{1}{4})\left(\frac{\xi}{r^2 f}\right)^{1/4}K_0(k_{0} \xi(\lambda)^{1/2} )d\lambda \nonumber\\
&- \mathcal{R}\left[\int^{\infty}_{0} \frac{ 2\lambda(\lambda^2 +\tfrac{1}{4})}{1 +e^{2\pi \lambda}} \left(\frac{\xi}{r^2 f}\right)^{1/4}K_0k_{0}( i \lambda) \xi^{1/2} )d\lambda\right] \nonumber\\
& =\mathcal{I}_1 +\mathcal{I}_2.
\end{align}
Here we have introduced a new integration variable $\lambda= l+1/2$ and $k^2_0(\lambda) =\lambda^2 +\hat{m}^2r_0^2 +1/12 \equiv \lambda^2 +k_{00}^2 $ with $k^2_0(i \lambda) =-\lambda^2 +k_{00}^2$.
\subparagraph{Evaluation of the integral $\mathcal{I}_1$\\}
We have 
\begin{align}
\mathcal{I}_1=A(r) \int^{\infty}_{0} \lambda(\lambda^2 -\tfrac{1}{4}) K_0(k_{0} (\lambda) \xi(r)^{1/2}) d\lambda
\end{align}
with $A(r)=\left(\frac{\xi}{r^2 f}\right)^{1/4}$. Using the relations $2k_0 dk = 2 \lambda d\lambda$ and $k_0^2(\lambda)= \lambda_0^2+k_{00}^2$, we obtain \cite{Watson}
\begin{align}
\mathcal{I}_{1}&=A(r) \int^{\infty}_{0} \lambda(\lambda^2 -\tfrac{1}{4})K_0(k(\lambda) \xi^{1/2})d\lambda\nonumber\\
&=A(r)\int^{\infty}_{k_0}k_0(\lambda)(k_0^2(\lambda) -k_{00}^2-\tfrac{1}{4})K_0(k_{0}(\lambda) \xi^{1/2}) dk \nonumber\\
&= A(r) \frac{8 k_{00}^2 \xi K_2(k_{00} \xi^{1/2}) -k_{00} \xi^{3/2}K_1(k_{00} \xi^{1/2}) }{4 \xi^2}.
\end{align}
\subparagraph{Evaluation of the integral $\mathcal{I}_2$\\}
We now consider the integral
\begin{align}
\mathcal{I}_2=- \mathcal{R}\left[\int^{\infty}_{0} \frac{ 2\lambda(\lambda^2 +\tfrac{1}{4})}{1 +e^{2\pi \lambda}}\left(\frac{\xi}{r^2 f}\right)^{1/4}K_0(k_{0}( i \lambda) \xi^{1/2} )d\lambda\right].
\end{align}
Due to the exponential factor in the denominator, this integral is absolutely convergent. Hence we can take the near horizon limit inside the sum to give 
\begin{align}
 \label{Eq:I2}
\mathcal{I}_2&=\frac{1}{\sqrt{\kappa r_0^2}}\mathcal{R}\left[\int^{\infty}_{0} \frac{ 2\lambda(\lambda^2 +\tfrac{1}{4})d\lambda}{1 +e^{2\pi \lambda}}\left\{ \frac{1}{2}\ln\left(\frac{k_{0}(i\lambda)^2\epsilon}{2 \kappa r_0^2} \right)+\gamma\right\} \right] \nonumber\\
&=\frac{1}{\sqrt{\kappa r_0^2}}\left\{\frac{34}{1920}\left(\gamma +\frac{1}{2}\ln\left(\frac{\epsilon}{2 \kappa r_0^2} \right)\right)\right.\nonumber\\
&\left. +\mathcal{R}\left[\int^{\infty}_{0} \frac{ 2\lambda(\lambda^2 +\tfrac{1}{4})}{1 +e^{2\pi \lambda}}\ln\left(k_{0}(i\lambda) \right) d\lambda\right]\right\} +O(\epsilon\ln(\epsilon))
\end{align}
where $\gamma$ is Euler's constant and we have used the integral formulas \cite{gradriz}
\begin{align}
\label{Eq:intform}
\int_{0}^{\infty}\frac{\lambda d\lambda}{1 +e^{2\pi \lambda}} =\frac{1}{48}; \quad \int_{0}^{\infty}\frac{\lambda^3d\lambda}{1 +e^{2\pi \lambda}} =\frac{7}{1920}.
\end{align}
Note that we truncate the series expansions at the level which contributes on the horizon.\\
The integral in Eq.~(\ref{Eq:I2}) has a branch point at $\lambda=k_{00}(k_0(i\lambda)=0)$, however the divergence  is logarithmic and therefore is integrable. We then may write
 \begin{align}
 \mathcal{R}\left[\int^{\infty}_{0} \frac{ 2\lambda(\lambda^2 +\tfrac{1}{4})}{1 +e^{2\pi \lambda}}\ln\left(k_{0}(i\lambda) \right)d\lambda \right]\nonumber\\
  = \int^{k_{00}}_{0} \frac{ 2\lambda(\lambda^2 +\tfrac{1}{4})}{1 +e^{2\pi \lambda}}\ln\left(k_{0}(i\lambda) \right)d\lambda \nonumber\\ +\int^{\infty}_{k_{00}} \frac{ 2\lambda(\lambda^2
 +\tfrac{1}{4})}{1 +e^{2\pi \lambda}}\ln\left(\hat{k}_{0}(i\lambda) \right)d\lambda
 \end{align}
 where $\hat{k}_0^2({i\lambda})=\lambda^2 -k_{00}^2$.

Hence, 
\begin{align}
\label{Eq:S1b}
&S_{1b}=\mathcal{I}_1 +\mathcal{I}_2 = \nonumber\\
 &A(r) \frac{8 k_{00}^2 \xi K_2�(k_{00} \xi^{1/2}) -k_{00} \xi^{3/2}K_1(k_{00} \xi^{1/2}) }{4 \xi^2}\nonumber\\
&+\frac{1}{\sqrt{\kappa r_0^2}}\left\{\frac{34}{1920}\left(\gamma +\frac{1}{2}\ln\left(\frac{\epsilon}{2 \kappa r_0^2} \right)\right)\right.\nonumber\\
&\left. +\int^{k_{00}}_{0} \frac{ 2\lambda(\lambda^2 +\tfrac{1}{4})}{1 +e^{2\pi \lambda}}\ln\left(k_{0}(i\lambda)\right)d\lambda \right.\nonumber\\
 &\left.+\int^{\infty}_{k_{00}} \frac{ 2\lambda(\lambda^2 +\tfrac{1}{4})}{1 +e^{2\pi \lambda}}\ln\left(\hat{k}_{0} (i \lambda) \right)d\lambda \right\} +O(\epsilon\ln(\epsilon)),
\end{align}
where we haven retained only those terms which contribute in the $\epsilon \to 0$ limit.
\paragraph{\bf Evaluation of the integral $S_{1a}$\\}
We calculate $S_{1a}$ in the following manner. Firstly we expand the summand in the small $\epsilon$ limit and isolate the divergence over $l$ by subtracting away the divergent terms (if any). We then express the sum over $l$ of these divergent terms as a geometric term which diverges in the $\epsilon \to 0$ limit and add this back on outside the sum. Finally, we apply the Riemann sum argument, developed in \cite{PhiSq}, to the original summand minus its divergent counterterm, to give the final value of the sum in the $\epsilon \to 0$ limit.

Taking the limit inside the sum gives
\begin{align}
S_{1a} = \frac{1}{\sqrt{\kappa rh^2}}\left\{(2l+1)\frac{l(l+1)}{4}[\ln(\nu_{0}) -\psi(\tfrac{1}{2}+\nu_0)]  \right\}\nonumber\\
+O(\epsilon)
\end{align}
and then, expanding each of the terms in the large $l$ limit,
we see that
   \begin{align}
(2l+1)\frac{l(l+1)}{4}[\ln(\nu_0) -\psi(\tfrac{1}{2}+\nu_0)]= -\frac{4\psi_0}{3 l}+O(l^{-2}).
\end{align}    
Using the identity derived in \cite{PhiSq}, 
\begin{equation*}
\ln(\eta -1) =-2\sum_{l=0}^{\infty} \frac{1}{(l+1)(\sqrt{\eta-1} +1)^{l+1}}+O(\sqrt{\eta-1}),
\end{equation*}
and letting $\eta =r/r_0$ allows us to express $S_{1a}$ in terms of a finite sum over $l$ plus a geometric term which diverges logarithmically as $\epsilon \to 0$:
 \begin{align}
& S_{1a}=    \sum_{l=0}^{\infty}\bigg\{\left.(2l+1)\frac{l(l+1)}{2}\right.\nonumber\\
&\left.\left. \bigg[\frac{1}{(r^2 f \xi)^{1/4}}\frac{\Gamma\left(\textstyle{\frac{1}{2} +\nu_0}\right)}{2^{3/2}( \psi_ 0)^{1/4}}W_{-\nu_0,0} \left(2\sqrt{\psi_0} \xi(r)\right)\right.\right.\nonumber\\
&\left.   \left.- A(r)K_{0}(k_0 \xi^{1/2})\right.\bigg]+A(r)\frac{4 \psi}{3} \frac{1}{(l+1)(\sqrt{\frac{\epsilon}{r_0}} +1)^{l+1}}\bigg\}\right.\nonumber\\
&+A(r)\frac{2 \psi}{3} \ln\left(\frac{\epsilon}{r_0}\right)+O(\sqrt{\epsilon}).
    \end{align}
 Denoting the summand as $F(l,\epsilon)$ we see the quantity we wish to calculate can be written as
     \begin{align}
     \label{Eq:S1riemann}
     \lim_{\epsilon \to 0}\left[\sum_{l=0}^{\infty}F(l,\epsilon)+A(r)\frac{2 \psi}{3} \ln\left(\frac{\epsilon}{r_0}\right)\right].
         \end{align}
Eq.~(\ref{Eq:S1riemann}) is now in a form which is amenable to the application of the Riemann sum argument. Using the methodology of  \cite{PhiSq} we expand $F(l,\epsilon)$ in the following manner:
       \begin{align}
       &F(l,\epsilon)=F_{0}(l) +F_1(x_l)\sqrt{\epsilon} +\Delta F(x_l,\epsilon)
       \end{align}
       where
       \begin{align}
       &F_0(l)=\lim_{\epsilon \to 0} F(l,\epsilon)\nonumber\\
&=\frac{1}{\sqrt{\kappa r_0^2}}\left\{(2l+1)\frac{l(l+1)}{4}[\ln(\nu_0) -\psi(\tfrac{1}{2}+\nu_0)]\right.\nonumber\\
&\left.+\frac{4 \psi_0}{3} \frac{1}{(l+1)}\right\},\nonumber\\
\label{defF1}
F_1(x_l)&=\lim_{\epsilon \to 0}\frac{1}{\sqrt{\epsilon}} F\bigg(\frac{x}{\sqrt{\epsilon}},\epsilon\bigg).
\end{align}     
Here $x_l=(l+1/2)\sqrt{\epsilon}$ and where recall from Eq.~(\ref{Eq:Q0def}) that $\nu_0=k_0^2/8\sqrt{\psi_0}$.
Unfortunately, in this case, it appears that we have to calculate the sum of $F_0(l)$ numerically. We do this, for a particular space-time, using the Levin $u$ transform (see \cite{Levrev} for a detailed discussion of this transform).

Now we turn our attention to $F_1(x_l)$.
Firstly, we expand $\xi$ in the near horizon limit
   \begin{align}
   \xi =\alpha^2 \epsilon +\beta \epsilon^2 +O(\epsilon^3);\quad \alpha= \sqrt{\frac{2}{\kappa r_0^2}} ;\nonumber\\ \beta =-\frac{2}{3}\left(\frac{8\kappa^2 +f_0'' r_0}{4\kappa^2 r_0^3}\right).
   \end{align}
   Then we may write  $ F\left(\displaystyle{\frac{x}{\sqrt{\epsilon}}},\epsilon\right)$ as
   \begin{align}
   \label{Eq:expanF1}
 \frac{x_l}{\sqrt{\epsilon}}\left(\frac{x_l^2}{\epsilon} -\frac{1}{4}\right)\left\{\left(\frac{1}{\sqrt{2 \epsilon}} +\frac{\beta}{2^{3/2} \alpha^2} \epsilon^{3/2}\right) \frac{\Gamma\left(\textstyle{\frac{1}{2} +\tilde{\nu}_0}\right)}{2^{3/2}( \psi_ 0)^{1/4}}\right.\nonumber\\
 \left. \times W_{-\tilde{\nu}_0,0}\left(2\sqrt{\psi_0} (\alpha^2 \epsilon +\beta \epsilon^2 )\right)\right.\nonumber\\
 \left.- \left(\frac{1}{\sqrt{\kappa r_0^2}} +\frac{\beta}{2^{1/2}\alpha}  x \right) K_0\left(8\sqrt{\psi_0}\tilde{\nu}_0\left(\alpha \epsilon^{1/2}+\frac{\beta}{2\alpha} \epsilon^{3/2}\right)\right)\right\}\nonumber\\
 +\left(\frac{1}{\sqrt{\kappa rh^2}} +\frac{\beta}{2^{1/2}\alpha}  x \right) \frac{4 \psi_0}{3} \frac{1}{(\tfrac{x_l}{\epsilon} +\tfrac{1}{2})(\sqrt{\frac{\epsilon}{r_0}} +1)^{\tfrac{x_l}{\epsilon} +1/2}} ,
   \end{align} 
where
\begin{align}
\label{Eq:nuhat}
\tilde{\nu}_0= \frac{x_l^2/\epsilon +\hat{m}^2 r_0^2+1/12}{8\sqrt{\psi_0}}.
\end{align}
Expanding the last term in Eq.~(\ref{Eq:expanF1}) about $\epsilon =0$ gives
\begin{align}
\label{Eq:expexpan}
\frac{4 \psi  e^{-\frac{x_l}{\sqrt{r_0}}}}{3 x_l r_0 \sqrt{\kappa }} \epsilon^{1/2} +O(\epsilon).
\end{align}
   We may also expand the second term in Eq.~(\ref{Eq:expanF1}) about $\epsilon =0$ to obtain
    \begin{align}
     \label{Eq:Besexpan}
&   \frac{x_l^3}{\sqrt{\kappa r_0^2}} \frac{K_0(\alpha x_{l})} {\epsilon^{3/2}} +\frac{1}{16 \sqrt[4]{2} \kappa ^{5/4} (r_0 \alpha )^{3/2}}\nonumber\\
&    \times\frac{1}{\epsilon^{1/2}}\left\{x_l \left(K_0(x_l \alpha ) \left(f''_0 r_0 x_l^2 \alpha ^2+4 \kappa  \left(\alpha ^2 \left(r_0+2
   x_l^2\right)\right.\right.\right.\right.\nonumber\\
 &  \left.\left.\left.\left.-r_0 x_l^2 \beta \right)\right)+2 r_0 x_l \alpha  \kappa  K_1(x_l \alpha ) \left((4 \hat{m}^2 r_0^2-1) \alpha
   ^2+4 x_l^2 \beta \right)\right) \right\}
   \nonumber\\
  & +G(x_l) \epsilon^{1/2},
     \end{align}
     where $G(x_l)$ is a rather long expression involving combinations of Bessel functions.\\
  We now require a similar expansion for the first term, which we would expect to cancel with the $\epsilon \to 0$ divergent  terms in the above expression, leaving us with a finite term that is integrable  over $x$ by the Riemann sum argument.\\
 To achieve this we consider the following integral representation for the Whittaker $W$ function:
 \begin{align}
    \label{Eq:WhitInt}
 \Gamma(\tfrac{1}{2} +\nu)W_{-\nu,0}(z)=\frac{2 \sqrt{z} e^{-z/2}}{\Gamma(\tfrac{1}{2} +\nu)}\int_{0}^{\infty} e^{-t}t^{\nu -1/2} K_{0}(2\sqrt{zt}) dt.
 \end{align}
We now apply Laplace's method, discussed in Appendix \ref{Ap:Laplace} for finding an asymptotic solution of this integral as $\nu \to \infty$. Application of this method leads to the following expansion for the first term in Eq.~(\ref{Eq:expanF1}): 
    \begin{align}
    \label{Eq:Whitfinalexpan}
&    \frac{x_l^3}{\sqrt{\kappa r_0^2}} \frac{K_0(\alpha x_{l})} {\epsilon^{3/2}} +\frac{1}{16 \sqrt[4]{2} \kappa ^{5/4} (r_0 \alpha )^{3/2}}\nonumber\\
&    \times\frac{1}{\epsilon^{1/2}}\left\{x_l \left(K_0(x_l \alpha ) \left(f''_0 r_0 x_l^2 \alpha ^2+4 \kappa  \left(\alpha ^2 \left(r_0+2
   x_l^2\right)\right.\right.\right.\right.\nonumber\\
 &  \left.\left.\left.\left.-r_0 x_l^2 \beta \right)\right)+2 r_0 x_l \alpha  \kappa  K_1(x_l \alpha ) \left((4 \hat{m}^2 r_0^2-1) \alpha
   ^2+4 x_l^2 \beta \right)\right) \right\}
   \nonumber\\
  & +\tilde{G}(x_l) \epsilon^{1/2} \end{align}
 which we may combine with the expansions (\ref{Eq:Besexpan}) and (\ref{Eq:expexpan}), achieving the result
     \begin{align}
& F\left(\frac{x_l}{\sqrt{\epsilon}},\epsilon\right)=-\frac{4\psi_0 \sqrt{\epsilon }}{3 r_0^5 \kappa ^2 x_l} \left[-r_0^3 \kappa ^{3/2}e^{-x_l/\sqrt{r_0}}\right.\nonumber\\
 &\left.+\sqrt{2} x_l
   \left(r_0^2 \kappa +x_l^2\right) K_1\left(x_l \alpha\right)+r_0 x_l^2 \sqrt{\kappa
   } K_0\left(x_l \alpha \right)\right].
  \end{align}
  One can readily numerically check that $\Delta F(x_l,\epsilon)$ is both $O(\epsilon)$ as $\epsilon \to 0$, and is a smooth integrable function. Therefore, we may apply the Riemann sum argument to give the relation
     \begin{align}
     S_{1a}=F_0(l) - \int_{0}^{\infty}\frac{4\psi_0 \sqrt{\epsilon }}{3 r_0^5 \kappa ^2 x_l} \left[-r_0^3 \kappa ^{3/2}e^{-x_l/\sqrt{r_0}}\right.\nonumber\\
 \left.+\sqrt{2} x_l
   \left(r_0^2 \kappa +x_l^2\right) K_1\left(x_l \alpha\right)+r_0 x_l^2 \sqrt{\kappa
   } K_0\left(x_l \alpha \right)\right]dx.
       \end{align}
       Using the identites contained in \cite{Watson}, we write down the value of the sum $S_{1a}$ in the horizon limit:
         \begin{align}
   S_{1a}=F_0(l) - \frac{\alpha ^3 \kappa  \psi_0 r_0  \left(\ln \left(r_0 \alpha
   ^2/4\right)-3\right)}{3 \sqrt{2}}\nonumber\\
   +A(r)\frac{2 \psi_0}{3} \ln\left(\frac{\epsilon}{r_0} \right)+O(\epsilon).
       \end{align} 
Together with Eq.~(\ref{Eq:S1b}) we finally arrive at an expression for the sum $S_1$ valid in the near horizon limit:
       \begin{align}
      \label{Eq:S1}
&S_1= \sum_{l=0}^{\infty}F_0(l) - \frac{\alpha ^3 \kappa  \psi_0  r_0\left(\ln \left(r_0 \alpha
   ^2/4\right)-3\right)}{3 \sqrt{2}} \nonumber\\
   &+A(r)\frac{2 \psi_0}{3} \ln\left(\frac{\epsilon}{r_0} \right)+\frac{34}{1920}\left(\gamma +\frac{1}{2}\ln(\epsilon)\right)\nonumber\\
&+A(r) \frac{8 k_{00}^2 \xi K_2(k_{00} \xi^{1/2}) -k_{00} \xi^{3/2}K_1(k_{00} \xi^{1/2}) }{4 \xi^2}
\nonumber\\
& +\int^{k_{00}}_{0} \frac{ 2\lambda(\lambda^2 +\tfrac{1}{4})}{1 +e^{2\pi \lambda}}\ln\left(k_{0}(i\lambda) \right)\nonumber\\
&+\int^{\infty}_{k_{00}} \frac{ 2\lambda(\lambda^2 +\tfrac{1}{4})}{1 +e^{2\pi \lambda}}\ln\left(\hat{k}_{0} (i \lambda) \right) +O(\epsilon).
     \end{align} 
 \subsection{Renormalized Values}
We are now in a position to substitute the expression (\ref{Eq:S1}) in the definition of $\{g^{\theta\theta'}G_{;\theta\theta'}$\}  Eq. (\ref{Eq:Gththp}) and expand in the near horizon limit. This procedure yields
  \begin{align}
&\{g^{\theta\theta'}G_{;\theta\theta'}\}=\frac{\kappa^2}{8\pi^2 \epsilon^2} - \frac{\kappa}{16 \pi^2 \epsilon} \left(\hat{m}^2 -\tfrac{1}{6} f_0'' +\tfrac{1}{3 r_0^2}\left(1+r_0 f_0'\right)\right)\nonumber\\
&+\frac{1}{2880\pi^2 r_0^4}\left\{ 1 +240\psi_0 -15 \hat{m}^2 r_0^2(2 +3 \hat{m}^2 r_0^2))\right\}\ln\left( \epsilon\right)\nonumber\\
& +O(1).
\end{align}
Comparing this expression with the renormalization subtraction terms (\ref{Eq:Gththpdiv}) and using the definition of $\psi_0$, it is straightforward to show that these divergences will cancel exactly with those contained in the subtraction terms, giving us the following renormalized value on the horizon:
         \begin{align}
         \label{Eq:Gththpren}
&[g^{\theta\theta'}G_{;\theta\theta'}]_{ren}= [g^{\theta\theta'}G_{;\theta\theta'}]_{numeric}+[g^{\theta\theta'}G_{;\theta\theta'}]_{analytic}
\end{align}
where
\begin{align}
&8\pi^2 r_0^4[g^{\theta\theta'}G_{;\theta\theta'}]_{numeric}
=\sum_{l=0}^{\infty} \left(\frac{4 \psi_0}{3} \frac{1}{(l+1)}\right.\nonumber\\
& \left.+  (2l+1)\frac{l(l+1)}{4}[\ln(\nu_0) -\psi(\tfrac{1}{2}+\nu_0)+2\beta^{W}_{0l}]\right)\nonumber\\
&+\int^{k_{00}}_{0} \frac{ 2\lambda(\lambda^2 +\tfrac{1}{4})}{1 +e^{2\pi \lambda}}\ln\left(k_{0}(i\lambda) \right)\nonumber\\
&+\int^{\infty}_{k_{00}} \frac{ 2\lambda(\lambda^2 +\tfrac{1}{4})}{1 +e^{2\pi \lambda}}\ln(\hat{k}_{0} (i \lambda)) 
\end{align}
\begin{align}
&8\pi^2 r_0^4[g^{\theta\theta'}G_{;\theta\theta'}]_{analytic}=\nonumber\\
&\frac{34}{1920}\left(\gamma -\frac{1}{2}\ln\left(2 \kappa r_0^2 \right)\right)
- \frac{2  \psi_0 r_0  \left(\ln \left(\frac{r_0 \alpha
   ^2}{4}\right)-3\right)}{3}\nonumber\\
&+\frac{1}{480} \left(r_0 \left(36 f'^2 r_0+2
   f'_0 \left(-8 f''_0 r_0^2+f_0'''
   r_0^3+15\right)-f^{''2}_0 r_0^3\right)\right.\nonumber\\
&\left.+30 k_{00}^2
   (4 f' r_0-2 \gamma +1)+30 \left(2
   k_{00}^4+k_{00}^2\right) \ln
   \left(\frac{k_{00}^2}{f' r_0^2}\right)\right.\nonumber\\
   &\left.+30 (3-4
   \gamma ) k_{00}^4\right)-\frac{2 \psi_0}{3} \ln\left(r_0\right) +\frac{F_{\theta \theta'}}{8\pi^2 r_0^4}.
 \end{align}
where we recall that $k_{00}=\sqrt{\hat{m}^2r_0^2 +1/12 }$ and $F_{\theta \theta'}$ is defined in Eq.~(\ref{Eq:Fthth}).\\
Repeating the procedure for the other derivatives leads to the following expressions on the horizon, again splitting each term into a numerical and an analytical component
\begin{align}
\label{Eq:GWren}
&8 \pi^2 r_0^2[G]_{numeric}=  \sum_{l=0}^{\infty} \frac{(2l+1)}{2}\left( \ln(\nu_0)-\psi(\tfrac{1}{2} +\nu_0)\right.\nonumber\\
 &\left.+2\beta^{W}_{0l}\right)-\int_0^{k_{00}} \frac{4 \lambda}{1+e^{2 \pi \lambda}} \ln(k_{0i \lambda}) d\lambda\nonumber\\
& -\int_{k_{00}}^{\infty} \frac{4 \lambda}{1+e^{2 \pi \lambda}} \ln(k_{0i \lambda}) d\lambda
\end{align}
 \begin{align}
&8 \pi^2 r_0^2[G]_{analytic}= \frac{1}{24} \left[12 k_{00}^2 \ln \left(\frac{k_{00}^2}{2
 r_0^2 \kappa }\right)+24 \gamma  k_{00}^2\right.\nonumber\\
  &\left.-12
   k_{00}^2-\ln\left(\frac{1}{r_0^2 \kappa }\right)-2 \gamma
   +\ln (2)\right] -\frac{1}{2} \hat{m}^2 \ln \left(\frac{\lambda}{\kappa
   }\right)+\frac{\kappa r_0 }{3 }
   \end{align}
\begin{align}
&8\pi^2 r_0^4[g^{tt'}G_{;tt'}]_{numeric}=\nonumber\\
&\sum_{l=0}^{\infty}\left(\frac{(2l+1)}{8}[2 k_1^2(\psi(1+\nu_1)-\ln(\nu_1) ) -8\sqrt{\psi_1}+8r_0^2\beta_{1l}]\right.\nonumber\\
&\left.+\frac{8 \psi_1}{3} \frac{1}{(l+1)} \right)
 \end{align}
 \begin{align}
&8\pi^2 r_0^4[g^{tt'}G_{;tt'}]_{analytic}=\nonumber\\
  &-\frac{4\psi_1  (\ln(2 r_0 \kappa )+1)}{3} -\frac{4 \psi_1}{3} \ln\left(r_0\right)\nonumber\\
 & + \frac{1}{288}\left.\bigg[r_0^2 \left(5 f_0''^2 r_0^2-4 r_0 \kappa  (4 f_0''+3 f_0''' r_0)-16 \kappa
   ^2\right)\right.\nonumber\\
 &\left.+24 k_{10}^2 r_0 (f_0'' r_0+2 \kappa ) +(54-72 \gamma ) k_{10}^4\right.\nonumber\\
&\left. -36 k_0^4 \left(2 \ln (k_{10})-\ln
   \left(2 r_0^2 \kappa \right)\right)\right.\bigg]\nonumber\\
&+\frac{1}{144}\left(-r_0 (f_0''r_0+2 \kappa )\right)+\frac{\kappa^2}{4 \pi^2 r_0^2 \sqrt{2}}P+\frac{F_{tt'}}{8\pi^2 r_0^4}
\nonumber\\
&+\frac{40k_{10}^2 -7}{5760}\left(-3 \ln \left(2r_0^2 \kappa \right)+(6 \gamma -3) \right)
 \end{align}
 \begin{align}
    &8\pi^2 r_0^4[ g^{tt}G_{;tt}]_{numeric}=\nonumber\\
    &\sum_{l=0}^{\infty}(2l+1)\left[\frac{f''_0 r_0^2 +4 f'_0 r_0}{24} \left(\psi(\tfrac{1}{2} +\nu_0)- \ln(\nu_0\right)\right.\nonumber\\
    &    \left.+\frac{k_0^2}{4}\left(\ln(\nu_0)-\psi(\tfrac{1}{2} +\nu_0)+\frac{4 \psi_0}{3l}\right)\right]\nonumber\\
    &  +\int^{k_{00}}_{0} \frac{ 2\lambda\ln(k_{0}(i\lambda))}{1 +e^{2\pi \lambda}}\left(\frac{r_0 (f''_0r_0 + 4 f'_0)}{6}-k_{0}(i\lambda)^2\right)d\lambda\nonumber\\
&  + \int_{k_{00}}^{\infty} \frac{ 2\lambda\ln(\hat{k}_{0} (i \lambda))}{1 +e^{2\pi \lambda}}\left(\frac{r_0 (f''_0r_0 + 4 f'_0)}{6}-\hat{k}_{0} (i \lambda)^2\right)d\lambda\nonumber\\
&+\frac{1}{2}\sum^{\infty}_{l=0}(2l+1) [l(l+1) +(m^2 +\xi R)r_0^2]\beta^{W}_{0l}              
  \end{align}
  \begin{align}
             &8\pi^2 r_0^4[ g^{tt}G_{;tt}]_{analytic}=-\frac{2\psi_0}{3} \ln(r_0)\nonumber\\
  &  +\frac{1}{480}\left[r_0^2 \left(f''^2 r_0^2-4 r_0 \kappa  (2 f''+31 f''' r_0) +16 \kappa^2\right)\right.\nonumber\\
  &   \left.
 +20 k_{00}^2 \left(3 k_{00}^2-r_0 (f'' r_0+8 \kappa )\right) \ln
   \left(\frac{k_{00}^2}{2 r_0^2 \kappa }\right)\right.\nonumber\\
   &\left.-20 k_{00}^2 r_0 ((2 \gamma -5) f'' r_0+8 (1+2
   \gamma ) \kappa )+30 (4 \gamma -3) k_{00}^4\right] \nonumber\\
& +\frac{1 }{288}\left[-\frac{f''_0 r_0^2 +4 f'_0 r_0}{6}\ln \left(2r_0^2 \kappa\right)+2 (\gamma
   -2) f'' r_0^2\right.\nonumber\\
  & \left.+16 (1+\gamma ) \kappa r_0\right]+ \frac{40 k_{00}^2 -7}{1920}\left( 1 -2 \gamma +\ln \left(2r_0^2 \kappa \right)\right)r_0\nonumber\\
  &+\frac{F_{tt}} {8\pi^2 r_0^4}.              
 \end{align}
 Note the coefficient of $\beta^{W}_{0l}$ in $[ g^{tt}G_{;tt}]_{numeric}$ comes from near horizon series expansion of $dp_{0l}/dr$.
It is also found that
\begin{align}
[ g^{tt'}G_{;tt'}]_{ren}=[ g^{rr'}G_{;rr'}]_{ren}\nonumber\\
[ g^{\theta\theta}G_{;\theta\theta}]_{ren}=-[ g^{\theta\theta'}G_{;\theta\theta'}]_{ren}
 \end{align}
 up to terms proportional to $R'(r)$. Of course, due to spherical symmetry we must have that 
 \begin{align}
 [ g^{\theta\theta}G_{;\theta\theta}]_{ren}=[ g^{\phi\phi}G_{;\phi\phi}]_{ren}\nonumber\\
[ g^{\theta\theta}G_{;\theta\theta}]_{ren}=[ g^{\phi\phi'}G_{;\phi\phi'}]_{ren}
 \end{align} \\
Finally we turn our attention to the  calculation of $[g^{rr}G_{;rr}]_{ren}$, whose construction follows immediately from the above expression. To see this we exploit  the fact that $[W(x,x'])$ (the coincidence limit of the regular part of the Hadamard expansion for $G_{E}$) satisfies the inhomogeneous wave equation \cite{BrownOttewill}
 \begin{align}
 (\Box -m^2 -\xi R)[W] = -6v_1,
 \end{align}
  where for a static spherically symmetric  Ricci-constant space-time $v_1$ is of the form
   \begin{align}
 v_1 &=\tfrac{1}{720} R_{abcd}R^{abcd} -\tfrac{1}{720} R_{ab}R^{ab} +\frac{\hat{m}^4}{8}\nonumber\\
 &=\frac{1}{1440 r^4}\left\{r^4 f''^2-8 f \left(r f'+1\right)+f' \left(8 r-4 r^3
   f''\right)\right.\nonumber\\
   &\left.+4 f^2+180 \hat{m}^4 r^4+4\right\}.
 \end{align}
Now since $[g^{rr}G_{E;rr}]_{ren}=[g^{rr}W_{;rr}]$ by definition, we have the result that
\begin{align}
\label{Eq:Gwave}
[g^{rr}G_{;rr}]_{ren}= -[g^{tt}G_{;tt}]_{ren}-[g^{\theta\theta}G_{;\theta\theta}]_{ren}\nonumber\\
- [g^{\phi\phi}G_{;\phi\phi}]_{ren} - (m^2 +\xi R)[G]_{ren} -6v_1
\end{align}
Inserting the expressions calulated for each term on the right hand side we arrive at the result that, on the horizon $r=r_0$,
 \begin{align}
[g^{rr}G_{;rr}]_{ren}= [g^{tt}G_{;tt}]_{ren}
\end{align}
again  up to terms proportional to $R'(r)$.
Finally we remark that equating the expression obtained here for $[G]_{ren}$ with that obtained in \cite{PhiSq} allows one to derive the following identity:
\begin{align}
&\int_0^{k_{00}} \frac{4 \lambda}{1+e^{2 \pi \lambda}} \ln(k_{0i \lambda}) d\lambda
 +\int_{k_{00}}^{\infty} \frac{4 \lambda}{1+e^{2 \pi \lambda}} \ln(k_{0i \lambda}) d\lambda\nonumber\\
&= k_{00}^2 \left(\ln(k_{00})-\tfrac{3}{2}\right)-\frac{d}{dx} \zeta\left(x,\textstyle{\frac{1}{2}} + i k_{00}\right) \bigg |_{x=-1} \nonumber\\
&-\frac{d}{dx} \zeta\left(x,\textstyle{\frac{1}{2}} - i k_{00}\right)\bigg|_{x=-1} +i k_{00}\ln\left(\frac{\Gamma\left(x,\textstyle{\frac{1}{2}} + i k_{00}\right)}{\Gamma\left(x,\textstyle{\frac{1}{2}} - ik_{00}\right)}\right).
\end{align} 

 \subsection{Renormalized Stress Tensor Components}
We begin this section by considering the behaviour of 
 \begin{align}
\langle \hat{T}_{r}^{~r}\rangle_{ren}-\langle \hat{T}_{t}^{~t}\rangle_{ren},
 \end{align}
near the horizons of a static spherically symmetric black hole space-time.
This expression, by Eq (\ref{Eq:Trenr}) and the fact that $R^{~t}_{t}=R^{~r}_{r}$ (hence $\mathcal{M}^{~t}_{t}=\mathcal{M}^{~r}_{r}$ by Eq.~(\ref{Eq:mcalug})), reduces to
  \begin{align}
&\langle \hat{T}_{r}^{~r}\rangle_{ren}-\langle \hat{T}_{t}^{~t}\rangle_{ren}=-2 \xi([g^{rr}G_{;rr}]_{ren}-[g^{tt}G_{;tt}]_{ren})\nonumber\\
&+(\tfrac{1}{2}-\xi)([g^{rr'}G_{;rr'}]_{ren}-[g^{tt'}G_{;tt'}]_{ren})
 \end{align} 
 As we have already shown that, on the horizon  $[g^{rr}G_{;rr}]_{ren}=[g^{tt}G_{;tt}]_{ren}$ and $[g^{rr}G_{;rr}]_{ren}=[g^{tt}G_{;tt}]_{ren}$,  we may therefore conclude that 
 \begin{align}
 \label{Eq:freefall}
\langle \hat{T}_{r}^{~r}\rangle_{ren}-\langle \hat{T}_{t}^{~t}\rangle_{ren}=0
 \end{align}
 on the horizon, $r=r_0$ of a general spherically symmetric space. \\

Now it is known, that in order for $\langle \hat{T}_{\mu}^{~\nu}\rangle_{ren}$ to be finite in a freely-falling frame on the past and future event horizons, the following conditions must be satisfied \cite{Babgen2d}:
\begin{align}
&1)~|\langle \hat{T}_{t}^{~t}\rangle_{ren}+\langle \hat{T}_{r}^{~r}\rangle_{ren}| < \infty.\nonumber\\
&2)~|\langle \hat{T}_{\theta}^{~\theta}\rangle_{ren}| < \infty.\nonumber\\
&3)~\frac{ |\langle \hat{T}_{t}^{~t}\rangle_{ren}-\langle \hat{T}_{r}^{~r}\rangle_{ren}|}{f} < \infty.
\end{align}
Its is straightforward to see that the first two of these conditions are satisfied. While Eq.~(\ref{Eq:freefall}) suggests that the third condition may be satisfied, the analysis of this paper does not allow us to prove this analytically, as we cannot rule out terms of the form $O(r-r_0)\ln(r-r_0)$ in the numerator. In order to complete this proof we require knowledge of  the behaviour of the derivatives of $\langle \hat{T}_{t}^{~t}\rangle_{ren}$ and $\langle \hat{T}_{r}^{~r}\rangle_{ren}$ as the horizon is approached. If they remain finite, then we may say that these quantities posses a Taylor series, at least to the first order, about the horizon, and hence we may then conclude that condition $3$ holds.

In Paper \textrm{II}, we calculate both $\langle \hat{T}_{r}^{~r}\rangle_{ren}$ and $\langle \hat{T}_{t}^{~t}\rangle_{ren}$ on the exterior region, excluding the immediate vicinity of the horizons, of a lukewarm black hole. Combining these numerical results with the horizon values obtained in this paper will allow us to numerically calculate these derivatives and thereby draw conclusions on the regularity of the equivalent of the Hartle-Hawking state for lukewarm black holes. We note here that the satisfaction of condition $3$ can be shown to be equivalent to requiring that $\langle \hat{T}_{\theta}^{~\theta}\rangle_{ren}$ possess a Taylor series about the horizon, by following the method of Morgan et al. \cite{Winstanleygen}. This will also be investigated in Paper \textrm{II}.

We may now write down expressions for the diagonal components of the stress tensor on a horizon of a spherically symmetric black hole space-time. These expressions are however quite unwieldy, fortunately with the approach adapted here we may easily unite them in terms of their sub-components.
  \begin{align}
&\langle \hat{T}_{r}^{~r}\rangle_{ren}=\langle \hat{T}_{t}^{~t}\rangle_{ren}= \nonumber\\
&2\xi( [g^{rr'}G_{rr'}]_{ren}+ [g^{rr}G_{rr}]_{ren})- [g^{\theta\theta'}G_{\theta\theta'}]_{ren}   \nonumber\\
&+\left\{\xi(R_{r}^{~r} -\tfrac{1}{2}R)-\tfrac{m^2}{2} \right\}[G]_{ren} +\frac{2 v_1}{8 \pi^2} +\mathcal{M}_{r}^{~r}
 \end{align}     
   \begin{align}
&\langle \hat{T}_{\theta}^{~\theta}\rangle_{ren}=\langle \hat{T}_{\phi}^{~\phi}\rangle_{ren}= \nonumber\\
&4\xi( [g^{rr'}G_{rr'}]_{ren}+ [g^{rr}G_{rr}]_{ren})- [g^{rr'}G_{rr'}]_{ren}\nonumber\\
&+\left\{\xi(R_{\theta}^{~\theta} -\tfrac{1}{2}R)-\tfrac{m^2}{2} \right\}[G]_{ren} +\frac{2 v_1}{8 \pi^2} +\mathcal{M}_{\theta}^{~\theta}
 \end{align}     
These expressions have the interesting consequence that for the minimally coupled case, one need only calculate $[g^{\theta\theta'}G_{\theta\theta'}]_{ren}$, $[g^{rr'}G_{rr'}]_{ren}$  and $[G]_{ren}$ in order to obtain the diagonal components of the stress tensor on a horizon, with the need for $[G]_{ren}$ disappearing in the massless case.
\section{Lukewarm Black Holes}
\label{sec:LW}
 We now specialize these results to the case of a lukewarm black hole. Lukewarm black holes are a special class of Reissner-Nordstrom-de Sitter space-times with (Euclidean) line element given by Eq.~(\ref{lee}) with metric function
\begin{equation}
\label{metric}
f (r) = 1 -\frac{2M}{r} +\frac{Q^2}{r^2} - \frac{\Lambda r^2}{3},
\end{equation}
where $M$, $Q$ are the mass and charge of the black hole respectively, and $\Lambda$ is the (positive) cosmological constant, with $Q=M$. 
For $4M<\sqrt{3/\Lambda}$ we have three distinct horizons, a black hole event horizon at $r=r_h$, an inner Cauchy horizon at $r=r_-$, and a cosmological horizon at $r=r_c$, where
\begin{subequations}
\begin{align}
r_-&=\frac{1}{2}\sqrt{{3}/{\Lambda}}\left(-1 +\sqrt{1 +4M\sqrt{{\Lambda}/{3}}}  \right).\\
r_h&=\frac{1}{2}\sqrt{{3}/{\Lambda}}\left(1 -\sqrt{1 -4M\sqrt{ {\Lambda}/{3}}}  \right).\\
r_c&=\frac{1}{2}\sqrt{{3}/{\Lambda}}\left(1 +\sqrt{1 -4M\sqrt{ {\Lambda}/{3}}}  \right).
\end{align}
\end{subequations}
The fourth  root of $f$ is negative and hence nonphysical.\\
 While the event horizon is formed by the gravitational potential of the black hole, the cosmological horizon is formed as a result of the expansion of the universe due to the cosmological constant \cite{GibHawk}. An observer located between the two horizons is causally isolated from the region within the event horizon, as well as from the region outside the cosmological horizon.\\
If, as the evidence seems to suggest, the universe possesses a cosmological constant \cite{Riess:1998}, then it is more natural to consider a black hole configuration which is asymptotically de Sitter than one which sits in an asymptotically flat universe. Also given that de Sitter space, in its natural vacuum, is awash with radiation to a static observer \cite{GibHawk}, it seems 
rather natural that a black hole in a de Sitter background would be most comfortable in a final configuration in which its event horizon is at the same temperature as the surrounding bath. This state of affairs is realized in the lukewarm case and so the study of such a  black hole configuration is well motived. In fact the lukewarm case has attracted interest recently, as evidenced in \cite{Winstanley:2007,PhiSq, Matyjasek:2011}.\\
We shall confine our attention to a single exterior region $r \in [r_h,r_c]$ which has a regular Euclidean section with topology $S^2 \times S^2$ \cite{Mellor}\\
In Tables~\ref{tab:Tththm} and \ref{tab:Tttm} we list the values of the pressure  $\langle \hat{T}_{\theta}^{~\theta}\rangle_{ren}$ and the energy density $-\langle \hat{T}_{t}^{~t}\rangle_{ren}$ for a range of valus of the mass of the field on both the event horizon and cosmological horizons respectively. We consider the case of a conformally coupled field of mass $m$ with $M=0.1L=Q$ with $L=3/\sqrt{\Lambda}$.\\
 As can be seen from the Tables, the energy density on the event horizon remains negative  as the mass of the field increases, while on the cosmological horizon it remains positive. The pressure remains positive on the event horizon, while its is mostly negative on the cosmogical horizon, as the field mass increases. \\
\begin{table}[htb]\centering
\begin{tabular}{| c | c | c | }\hline
 & & \\
 $m$ &  Event & Cosmological \\ \hline
 & & \\
 $0 $ & $0.3166133$ & $-0.0001567$ \\\
  & & \\\
 $\frac{1}{4} L$ & $0.31458112$ & $0.0016206$ \\
 & &\\
  $\frac{1}{2}  L$ &  $0.3033364$ & $-0.0008350$ \\
 & & \\
  $\frac{3}{4}  L $&  $0.2920777$ & $-0.0013648$ \\ 
& & \\
 $1 L$ &  $0.2829577$& $-0.0044883$  \\ 
 & & \\
\hline
 \end{tabular}
\caption{ $\langle \hat{T}_{\theta}^{~\theta}\rangle_{ren}$ on both horizons for various field masses.}
\label{tab:Tththm}
\end{table}
\begin{table}[htb]\centering
\begin{tabular}{| c | c | c | }\hline
 & & \\
 $m$ &  Event & Cosmological \\ \hline
 & & \\
 $0 $ & $-0.0058418$ & $0.0000407$ \\ 
  & & \\\
 $\frac{1}{4} L$ & $-0.0062360$ & $6.133531\times 10^{-6}$ \\
 & &\\
  $\frac{1}{2}  L$ &  $-0.0139295$ & $0.0007781$ \\
 & & \\
  $\frac{3}{4}  L $&  $-0.0171130$ & $0.0019650$ \\ 
& & \\
 $1 L$ &  $-0.0117374$& $0.0053876$  \\ 
 & & \\
\hline
 \end{tabular}
\caption{Energy density on both horizons for various field masses.}
\label{tab:Tttm}
\end{table}
To interpret this result we consider the semi-classical field equations:
\begin{align}
G_{\mu \nu} +\Lambda g_{\mu \nu} = \langle \hat{T}_{\mu \nu }\rangle_{ren}.
\end{align}
If we take the $\Lambda g_{\mu \nu} $ over to the right hand side and consider it to be a classical stress tensor contribution due to the cosmological constant. We see that, since the metric has signature $(-,+,+,+)$ and $\Lambda$ is positive, the energy density of this ``stress tensor"  and the pressure are both negative. One may say that it is this negative pressure that leads to inflation. We may conclude that incorporating the `one loop'  quantum effects causes an increase in energy density on the cosmological horizon while it adds to the negative pressure driving inflation for the massless conformally coupled case. Similar conclusions for different values of the mass may be drawn from the Tables ~\ref{tab:Tththm} and \ref{tab:Tttm}.

\section{Conclusions}
\label{sec:Conclusions}
The key question in study of quantum field theory on black hole spacetimes with multiple horizons, including the important class of  lukewarm black hole configurations, is whether it is possible,
when the temperature of the two horizons is equal, to define an equivalent of the Hartle-Hawking state on Schwarzschild space-time 
which is regular on both horizons?
In this paper, we have made major progress towards answering this question in the affirmative by using  the Hadamard renormalization procedure to calculate expressions for the components of the renormalized stress tensor $\langle \hat{T}_{\mu}^{\nu}\rangle_{ren}$, on the horizons of a spherically symmetric black hole spacetime. 
 Making use of uniform approximations to the solutions of the radial equations, we were able to calculate the value of $\langle \hat{T}_{\mu \nu}\rangle_{ren}$ for a quantum scalar field with arbitrary mass and coupling to the geometry and which is in a Hartle-Hawking state, on the horizons of a general, spherically symmetric black hole spacetime, which is not necessarily asymptotically flat. In particular we demonstrated, 
two of the three necessary conditions for regularity: finiteness of $\langle\hat{T}_{t}{}^{t} + \hat{T}_{r}{}^{r}\rangle_{ren}$ and of $\langle\hat{T}_{\theta}{}^{\theta}\rangle_{ren}$.

We also made progress in proving the third condition, namely, that the $|\langle\hat{T}_{t}{}^{t}  - \hat{T}_{r}{}^{r}\rangle_{ren}|/(r-r_h)$ be finite by proving that $\langle \hat{T}_{r}^{~r}\rangle_{ren}$ and $\langle \hat{T}_{t}^{~t}\rangle_{ren}$ components are equal on the two horizons of our
spacetime region. Therefore the regularity of the Hartle-Hawking state depends entirely on this difference of  components possessing a Taylor series to the first order about the horizon. 
To address this question, and to find the behaviour of the renormalized stress tensor away from the horizon, requires a different
numerical approach which we will present in Paper~II.
To underline the different nature of these calculation we note that to calculate quantities on the horizon it is most convenient to separate in a radial direction as only the lowest two frequency modes are required there for the Hartle-Hawking vacuum. By contrast, away from the horizon
temporal separation, which is not posssible on the horizon since the separation becomes null, is most convenient, all modes contribute
and we must resort to a numerical calculation.  
In Paper~II, we will complete the our `proof' of finiteness by providing compelling numerical evidence that $\langle\hat{T}_{t}{}^{t} \rangle_{ren}$
and $\langle \hat{T}_{r}{}^{r}\rangle_{ren}$ do indeed possess finite first derivatives at the horizons.

\acknowledgments
The work of CB is supported by the Irish Research Council for Science, Engineering and Technology, funded by the National Development Plan.

\appendix
\section{ $\beta^{W}_{0l}$ and $\beta^{W}_{1l}$}
 \label{Ap:Beta}
 In this Appendix we give expressions for $\beta^{W}_{0l}$ and $\beta^{W}_{1l}$ for the lukewarm black hole configuration. These calculations follow along the lines of Appendix B in \cite{PhiSq}, yielding the results, for the event horizon:
  \begin{align}
&\beta^{W}_{1l}=   \int^{r_c}_{r_h} I(r) \left(\frac{2p_{1l}'(r')}{p_{1l}(r')^3} -\frac{\hat{a}}{(r-r_h)^2}\right) dr + I_2(r_c)\nonumber\\
& -\frac{\hat{b}}{4(\kappa r_h^2)} 
+\frac{1}{2\kappa r_c^2}\left(b + \frac{a}{r_c-r_h}\right)\ln(r_c-r_h)\nonumber\\
&+ \frac{1}{2\kappa_i r_i^2}\left( \frac{a}{r_h-r_i}-b\right)\ln(r_h-r_i)+
 I_2(r_c) \nonumber\\
 &+ \frac{1}{2\kappa_n r_n^2}\left( \frac{a}{r_h-r_n}-b\right)\ln(r_h-r_n)\nonumber\\
&+ \frac{1}{48 (\kappa rh^2)^{3/2}}\left.\bigg[12(2 \gamma -1)k_1^2 -r_h(4 f'_h +f''_h)  \right.\nonumber\\
&\left.-48\sqrt{\psi_1}+12 k^2 \left(\ln\left(\frac{4 \sqrt{\psi_1}}{\kappa r_h^2}\right)+\psi(1 +\nu_1)\right)\bigg]\right.
\end{align}
which can be shown, numerically, to exhibit a $l^{-4}$ behaviour for large $l$. 
\begin{align}
\label{Eq:betaW0}
\beta^{W}_{0l} &= \int^{r_c}_{r_h} \frac{2 p'_l(r')}{p_l^3(r')}I(r') dr' +\frac{1}{4}\ln(\psi_1) +\gamma+\ln(2a_0)  \nonumber\\
&+\kappa r_h^2\bigg[\frac{1}{\kappa_c r_c^2} \ln(r_c-r_h)-\frac{1}{2\kappa_ir_i^2} \ln(r_h-r_i)\nonumber\\
&- \frac{1}{2\kappa_n r_n^2} \ln(r_h-r_n) \bigg]
+\frac{1}{2}\psi(\textstyle{\frac{1}{2} -\nu_0})
\end{align}
which behaves like $l^{-6}$ for large $l$.
Here
\begin{align}
&a =\frac{1}{\sqrt{\kappa r_h^2}}; \quad \hat{a}=\frac{1}{a^2} \nonumber\\
&b= \frac{a^3}{4 }\left(l(l+1) +(m^2 +\xi R)r_h^2 -\tfrac{r_h}{2}(2 f'_h +f''_h r_h)\right)\nonumber\\
& \hat{b}=\frac{1}{b^2}.\nonumber\\
&I(r)= \frac{1}{\kappa_h r_h^2} \ln(r-r_h)+\frac{1}{2\kappa_ir_i^2} \ln(r-r_i)\nonumber\\
&+\frac{1}{2\kappa_n r_n^2} \ln(r-r_n)-\frac{1}{2\kappa_c r_c^2} \ln(r_c-r).\nonumber\\
&I_2(r)=\int I(r)\frac{\hat{a}}{(r-r_h)^2} dr
\end{align}

Finally we note that all the relevant sums over $l$ are performed using the Levin $u$ transformation

\section{Laplace's Method}
\label{Ap:Laplace}
In this discussion we draw heavily from Murray \cite{Murray}.
Laplace's method is a technique for obtaining an asymptotic expansion of an integral of the form
 \begin{align}
 \label{Eq:Laplace}
 I(x) =\int^{a}_{-b} g(t) e^{h(t) x},
  \end{align}
 as $x\to \infty$.
 The main concept  behind the method is that if the function $h(t)$ has its maximum at $0$, and $g(0)\neq0$, then the dominant contribution to the large $x$ asymptotic expansion of $I(x)$ will come from the immediate neighbourhood of $0$. To calculate an asymptotic expansion for $I(x)$ as $x\to \infty$ it is appropriate to introduce a new variable, $s$, defined by the relation
 \begin{align}
 \label{Eq:strans}
 h(t)- h(0)=-s^2.
 \end{align}
 Hence, the exponential term in Eq. (\ref{Eq:Laplace}) becomes
  \begin{align}
e^{h(t) x}=e^{h(0) x}e^{-x s^2},
 \end{align}
 and so  Eq. (\ref{Eq:Laplace}) is of the form
  \begin{align}
  \label{Eq:murraylap}
 I(x) =\int^{A}_{-B} \phi(s) e^{-x s^2 }ds.
   \end{align}
By Watson's lemma, an asymptotic power series as $x\to \infty$ for Eq.~(\ref{Eq:murraylap}) may be obtained \cite{Murray}. 
  \begin{widetext}
 Following the arguments of \cite{Murray} one can obtain an asymptotic expansion of Eq. (\ref{Eq:Laplace}) as $x \to \infty$ given by:
          \begin{align}
     \label{Eq:Intexpan}
     I(x) \approx e^{x h(0)}\sqrt{\frac{2 \pi}{-h''(0) x}}\left\{g(0) +\frac{1}{24
   h''(0)^3 x} \left[-12 g''(0) h''(0)^2+12 h'''(0) g'(0) h''(0)+g(0) \left(3 h^{(4)}(0) h''(0)-5 h'''(0)^2\right)\right]\right.\nonumber\\
   \left.+\frac{1}{1152 h''(0)^6 x^2} \left[144 g^{(4)}(0) h''(0)^4-210 h'''(0)^2 h''(0) \left(4 h'''(0) g'(0)+3 g(0) h^{(4)}(0)\right)\right.\right.\nonumber\\
   \left.\left.+21 h''(0)^2
   \left(40 h'''(0)^2 g''(0)+5 h^{(4)}(0) \left(8 h'''(0) g'(0)+g(0) h^{(4)}(0)\right)+8 g(0) h^{(4)}(0)
   h'''(0)\right)\right.\right.\nonumber\\
\left.\left.-24 h''(0)^3 \left(20 g'''(0) h'''(0)+15 h^{(4)}(0) g''(0)+6 h^{(4)}(0) g'(0)+g(0)
   h^{(5)}(0)\right)+385 g(0) h'''(0)^4\right]+O(x^{-3})\right\}.
       \end{align}    
       Higher order coefficients can be found in \cite{Dingle}.\\
  Now, we shift our focus back to the integral in Eq.~(\ref{Eq:WhitInt}). If we introduce a new variable $\tau$ defined by $t=\nu(\tau+1) $, the integral takes the form
 \begin{align}
    \label{Eq:WhitIntTrans}
 \Gamma(\tfrac{1}{2} +\nu)W_{-\nu,0,z}=\frac{2 \sqrt{z} e^{-z/2}}{\Gamma(\tfrac{1}{2} +\nu)}\nu^{\nu +\tfrac{1}{2}}\int_{-1}^{\infty}  e^{\nu(-(\tau+1) +\ln(\tau+1))}(\tau+1)^{ -1/2} K_{0}\left(2\sqrt{z(\tau+1) \nu}\right) d\tau.
 \end{align}
     
Setting $\nu =\tilde{\nu}_0$ and considering the limit $\epsilon \to 0$ allows us to observe, by Eq.~(\ref{Eq:nuhat}),  that we require an asymptotic expansion of the above integral for $\nu \to \infty$. In other words, we may apply the expansion (\ref{Eq:Intexpan}) to the integral in Eq.~(\ref{Eq:WhitIntTrans}). Examining this integral we see that, in the notation of Eq.(\ref{Eq:Laplace})
 \begin{align}
 g(\tau) = (\tau+1)^{ -1/2} K_{0}\left(2\sqrt{z(\tau+1) \nu}\right); \quad h(\tau)=-(\tau+1) +\ln(\tau+1).
 \end{align}
 It is clear that $h(\tau)$ achieves a maximum value at $\tau=0$
 and so we have the following asymptotic expansion for the integral in Eq.(\ref{Eq:WhitIntTrans}):
  \begin{align}
  \label{Eq:Whitexpan}
& \int_{0}^{\infty}  e^{\nu(-\tau +\ln(\tau))}\tau^{ -1/2} K_{0}\left(2\sqrt{z\tau \nu}\right) \nu dt \approx   \sqrt{\frac{2\pi}{\nu} }e^{-\nu}\left\{K_{0}(2\sqrt{z \nu}) +\frac{1}{24 \nu}(12\nu z-1)K_{0}(2\sqrt{z\tau z})\right.\nonumber\\
& \left. \frac{1}{1152 \nu^2}\left[ (72\nu z(2\nu z-1)+1) K_{0}\left(2\sqrt{z\tau \nu}\right) -48\sqrt{\nu z}(1+2 z \nu)  K_{1}(2\sqrt{z\tau z})\right]\right\}.
  \end{align}
  Of course, in this case the function $g(\tau)$ is actually a function of both $\tau$ and $\nu$; however  we are interested only in taking the near horizon limit ($z \to 0$), and in this limit the product $\nu z$ tends to a constant, so we may treat it as so.

  Next, we consider the integral representation \cite{gradriz}
    \begin{align}
    \Gamma(\nu+\tfrac{1}{2})=\int_{0}^{\infty} e^{-t}t^{\nu-\tfrac{1}{2}}dt.
      \end{align}
      Applying Laplace's method to this integral, we obtain the well known large $\nu$ asymptotic expansion of $ \Gamma(\nu+\tfrac{1}{2})$:
 \begin{align}
\Gamma(\nu+\tfrac{1}{2})\approx \nu^{\nu} \sqrt{2\pi}e^{-\nu}\left\{1 -\frac{1}{24 \nu} +\frac{1}{1152 \nu^2} +O(\nu^{-3})\right\}.
  \end{align}
  Combining this expression with Eqs. (\ref{Eq:Whitexpan}) and (\ref{Eq:WhitInt}) yields the result
   \begin{align}
  & \Gamma(\tfrac{1}{2} +\nu)W_{-\nu,0}(z)\approx 2 \sqrt{z} e^{-z/2}\left[1 +\frac{1}{24\nu}+\frac{1}{1552 \nu^2}\right]\left\{K_{0}(2\sqrt{z \nu}) +\frac{1}{24 \nu}(12\nu z-1)K_{0}(2\sqrt{z\tau z})\right.\nonumber\\
& \left. \frac{1}{1152 \nu^2}\left[ (72\nu z\{2\nu z-1\}+1) K_{0}(2\sqrt{z\tau z}) -48\sqrt{\nu z}(1+2 z \nu)  K_{1}(2\sqrt{z\tau z})\right]\right\}.
 \end{align}
  \end{widetext}
Making the replacements
     \begin{align}
     z=2 \sqrt{\psi_0} \xi;\quad \xi\approx 2\sqrt{\psi_0} (\alpha^2 \epsilon +\beta \epsilon^2 ) ; \quad \nu = \tilde{\nu}_0.
  \end{align}
  we obtain the expansion (\ref{Eq:Whitfinalexpan}).

 It should be noted here that for some parameter sets $\psi_0$ is negative and so $\nu_0$ is purely imaginary; in this case Laplace's method is not applicable, and the method of stationary phase must be used instead. In this case we consider integrals of the form
 \begin{align}
 \int_{a}^{b} g(t) e^{i x h(t)}dt ,
 \end{align}
   as $x \to \infty$ ($x$ real), and assume that $h(t)$ has a turning point at $t=c$, $c \in (a,b)$. The term $e^{i x h(t)}$ is now purely oscillatory, and as $x$ increases these oscillations become more and more rapid, except at the turning point $c$. Near $c$, the phase $x h(t)$ is approximately constant  (or stationary) as $x \to \infty$. One can expand the integrand about this point, in a analogous manner to the analysis of Laplace's method, leading to a result similar to Eq. (\ref{Eq:Intexpan}) to first order. To achieve higher order approximations we require the method of steepest descent (see \cite{Murray} for details). Applying this method to Eq.~(\ref{Eq:WhitIntTrans}) when $\nu$ is purely imaginary leads precisely to the result (\ref{Eq:Whitfinalexpan}).


\begin{thebibliography}{40}
\expandafter\ifx\csname natexlab\endcsname\relax\def\natexlab#1{#1}\fi
\expandafter\ifx\csname bibnamefont\endcsname\relax
  \def\bibnamefont#1{#1}\fi
\expandafter\ifx\csname bibfnamefont\endcsname\relax
  \def\bibfnamefont#1{#1}\fi
\expandafter\ifx\csname citenamefont\endcsname\relax
  \def\citenamefont#1{#1}\fi
\expandafter\ifx\csname url\endcsname\relax
  \def\url#1{\texttt{#1}}\fi
\expandafter\ifx\csname urlprefix\endcsname\relax\def\urlprefix{URL }\fi
\providecommand{\bibinfo}[2]{#2}
\providecommand{\eprint}[2][]{\url{#2}}

\bibitem[{\citenamefont{Candelas}(1980)}]{Candelas}
\bibinfo{author}{\bibfnamefont{P.}~\bibnamefont{Candelas}},
  \bibinfo{journal}{Phys.\ Rev.\ D} \textbf{\bibinfo{volume}{21}},
  \bibinfo{pages}{2185} (\bibinfo{year}{1980}).

\bibitem[{\citenamefont{Howard and Candelas}(1984)}]{Howard:1984qp}
\bibinfo{author}{\bibfnamefont{K.~W.} \bibnamefont{Howard}} \bibnamefont{and}
  \bibinfo{author}{\bibfnamefont{P.}~\bibnamefont{Candelas}},
  \bibinfo{journal}{Phys.\ Rev.\ Lett.} \textbf{\bibinfo{volume}{53}},
  \bibinfo{pages}{403} (\bibinfo{year}{1984}).

\bibitem[{\citenamefont{Howard}(1984)}]{Howard:1985yg}
\bibinfo{author}{\bibfnamefont{K.~W.} \bibnamefont{Howard}},
  \bibinfo{journal}{Phys.\ Rev.\ D} \textbf{\bibinfo{volume}{30}},
  \bibinfo{pages}{2532} (\bibinfo{year}{1984}).

\bibitem[{\citenamefont{Anderson}(1989)}]{Anderson:1989vg}
\bibinfo{author}{\bibfnamefont{P.~R.} \bibnamefont{Anderson}},
  \bibinfo{journal}{Phys.\ Rev.\ D} \textbf{\bibinfo{volume}{39}},
  \bibinfo{pages}{3785} (\bibinfo{year}{1989}).

\bibitem[{\citenamefont{Anderson et~al.}(1995)\citenamefont{Anderson, Hiscock,
  and Samuel}}]{Anderson:1994hg}
\bibinfo{author}{\bibfnamefont{P.~R.} \bibnamefont{Anderson}},
  \bibinfo{author}{\bibfnamefont{W.~A.} \bibnamefont{Hiscock}},
  \bibnamefont{and} \bibinfo{author}{\bibfnamefont{D.~A.}
  \bibnamefont{Samuel}}, \bibinfo{journal}{Phys.\ Rev.\ D}
  \textbf{\bibinfo{volume}{51}}, \bibinfo{pages}{4337} (\bibinfo{year}{1995}).

\bibitem[{\citenamefont{DeBenedictis}(1999)}]{DeBenedictis}
\bibinfo{author}{\bibfnamefont{A.}~\bibnamefont{DeBenedictis}},
  \bibinfo{journal}{Gen. Rel. and Grav.} \textbf{\bibinfo{volume}{31}},
  \bibinfo{pages}{1549} (\bibinfo{year}{1999}), ISSN \bibinfo{issn}{0001-7701},
  \bibinfo{note}{10.1023/A:1026734521256},
  \urlprefix\url{http://dx.doi.org/10.1023/A:1026734521256}.

\bibitem[{\citenamefont{Elster}(1984)}]{Elster}
\bibinfo{author}{\bibfnamefont{T.}~\bibnamefont{Elster}},
  \bibinfo{journal}{Class. and Quan. Grav.} \textbf{\bibinfo{volume}{1}},
  \bibinfo{pages}{43} (\bibinfo{year}{1984}),
  \urlprefix\url{http://stacks.iop.org/0264-9381/1/i=1/a=007}.

\bibitem[{\citenamefont{Fawcett}(1983)}]{Fawcett:1983dk}
\bibinfo{author}{\bibfnamefont{M.~S.} \bibnamefont{Fawcett}},
  \bibinfo{journal}{Communications in Mathematical Physics}
  \textbf{\bibinfo{volume}{89}}, \bibinfo{pages}{103} (\bibinfo{year}{1983}).

\bibitem[{\citenamefont{Frolov and Zel'nikov}(1982)}]{FrolovZel1}
\bibinfo{author}{\bibfnamefont{V.~P.} \bibnamefont{Frolov}} \bibnamefont{and}
  \bibinfo{author}{\bibfnamefont{A.~I.} \bibnamefont{Zel'nikov}},
  \bibinfo{journal}{Physics Letters B} \textbf{\bibinfo{volume}{115}},
  \bibinfo{pages}{372 } (\bibinfo{year}{1982}).

\bibitem[{\citenamefont{Frolov and Zel'nikov}(1984)}]{FrolovZel2}
\bibinfo{author}{\bibfnamefont{V.~P.} \bibnamefont{Frolov}} \bibnamefont{and}
  \bibinfo{author}{\bibfnamefont{A.~I.} \bibnamefont{Zel'nikov}},
  \bibinfo{journal}{Phys. Rev. D} \textbf{\bibinfo{volume}{29}},
  \bibinfo{pages}{1057} (\bibinfo{year}{1984}).

\bibitem[{\citenamefont{Groves et~al.}(2002)\citenamefont{Groves, Anderson, and
  Carlson}}]{AndersonGroves}
\bibinfo{author}{\bibfnamefont{P.~B.} \bibnamefont{Groves}},
  \bibinfo{author}{\bibfnamefont{P.~R.} \bibnamefont{Anderson}},
  \bibnamefont{and} \bibinfo{author}{\bibfnamefont{E.~D.}
  \bibnamefont{Carlson}}, \bibinfo{journal}{Phys. Rev. D}
  \textbf{\bibinfo{volume}{66}}, \bibinfo{pages}{124017}
  (\bibinfo{year}{2002}).

\bibitem[{\citenamefont{Jensen et~al.}(1995)\citenamefont{Jensen, Mc.~Laughlin,
  and Ottewill}}]{JensenOttewill:95}
\bibinfo{author}{\bibfnamefont{B.~P.} \bibnamefont{Jensen}},
  \bibinfo{author}{\bibfnamefont{J.~G.} \bibnamefont{Mc.~Laughlin}},
  \bibnamefont{and} \bibinfo{author}{\bibfnamefont{A.~C.}
  \bibnamefont{Ottewill}}, \bibinfo{journal}{Phys. Rev. D}
  \textbf{\bibinfo{volume}{51}}, \bibinfo{pages}{5676} (\bibinfo{year}{1995}).

\bibitem[{\citenamefont{Christensen}(1976)}]{Christensen:1976vb}
\bibinfo{author}{\bibfnamefont{S.~M.} \bibnamefont{Christensen}},
  \bibinfo{journal}{Phys.\ Rev.\ D} \textbf{\bibinfo{volume}{14}},
  \bibinfo{pages}{2490} (\bibinfo{year}{1976}).

\bibitem[{\citenamefont{Brown and Ottewill}(1986)}]{BrownOttewill}
\bibinfo{author}{\bibfnamefont{M.~R.} \bibnamefont{Brown}} \bibnamefont{and}
  \bibinfo{author}{\bibfnamefont{A.~C.} \bibnamefont{Ottewill}},
  \bibinfo{journal}{Phys. Rev. D} \textbf{\bibinfo{volume}{34}},
  \bibinfo{pages}{1776} (\bibinfo{year}{1986}).

\bibitem[{\citenamefont{Wald}(1977)}]{Wald}
\bibinfo{author}{\bibfnamefont{R.~M.} \bibnamefont{Wald}},
  \bibinfo{journal}{Comms. in Math. Phys.} \textbf{\bibinfo{volume}{54}},
  \bibinfo{pages}{1} (\bibinfo{year}{1977}), ISSN \bibinfo{issn}{0010-3616},
  \bibinfo{note}{10.1007/BF01609833},
  \urlprefix\url{http://dx.doi.org/10.1007/BF01609833}.

\bibitem[{\citenamefont{Breen and Ottewill}(2010)}]{PhiSq}
\bibinfo{author}{\bibfnamefont{C.}~\bibnamefont{Breen}} \bibnamefont{and}
  \bibinfo{author}{\bibfnamefont{A.~C.} \bibnamefont{Ottewill}},
  \bibinfo{journal}{Phys. Rev. D} \textbf{\bibinfo{volume}{82}},
  \bibinfo{pages}{084019} (\bibinfo{year}{2010}),
  \urlprefix\url{http://link.aps.org/doi/10.1103/PhysRevD.82.084019}.

\bibitem[{\citenamefont{Winstanley and Young}(2008)}]{Winstanley:2007}
\bibinfo{author}{\bibfnamefont{E.}~\bibnamefont{Winstanley}} \bibnamefont{and}
  \bibinfo{author}{\bibfnamefont{P.~M.} \bibnamefont{Young}},
  \bibinfo{journal}{Phys. Rev. D} \textbf{\bibinfo{volume}{77}}
  (\bibinfo{year}{2008}).

\bibitem[{\citenamefont{Misner et~al.}(1973)\citenamefont{Misner, Thorne, and
  Wheeler}}]{MTW}
\bibinfo{author}{\bibfnamefont{C.~W.} \bibnamefont{Misner}},
  \bibinfo{author}{\bibfnamefont{K.~S.} \bibnamefont{Thorne}},
  \bibnamefont{and} \bibinfo{author}{\bibfnamefont{J.~A.}
  \bibnamefont{Wheeler}}, \emph{\bibinfo{title}{Gravation}}
  (\bibinfo{publisher}{W. H. Freeman and Company, San Francisco},
  \bibinfo{year}{1973}).

\bibitem[{\citenamefont{Birrell and Davies}(1982)}]{BD}
\bibinfo{author}{\bibfnamefont{N.~D.} \bibnamefont{Birrell}} \bibnamefont{and}
  \bibinfo{author}{\bibfnamefont{P.~C.~W.} \bibnamefont{Davies}},
  \emph{\bibinfo{title}{Quantum Fields in Curved Space}}
  (\bibinfo{publisher}{Cambridge Monographs on Mathematical Physics},
  \bibinfo{year}{1982}).

\bibitem[{\citenamefont{Garabedian}(1964)}]{Garabedian}
\bibinfo{author}{\bibfnamefont{P.}~\bibnamefont{Garabedian}},
  \emph{\bibinfo{title}{Partial Differential Equations}}
  (\bibinfo{publisher}{Wiley, New York}, \bibinfo{year}{1964}).

\bibitem[{\citenamefont{J.Hadamard}(1923)}]{Hadamard}
\bibinfo{author}{\bibnamefont{J.Hadamard}}, \emph{\bibinfo{title}{Lectures on
  Cauchy's Problem in Linear Partial Differential Equations}}
  (\bibinfo{publisher}{Yale University Press, New Haven},
  \bibinfo{year}{1923}).

\bibitem[{\citenamefont{DeWitt and Brehme}(1960)}]{DeWittBrehme}
\bibinfo{author}{\bibfnamefont{B.~S.} \bibnamefont{DeWitt}} \bibnamefont{and}
  \bibinfo{author}{\bibfnamefont{R.~W.} \bibnamefont{Brehme}},
  \bibinfo{journal}{Ann. Phys (N.Y)} \textbf{\bibinfo{volume}{9}}
  (\bibinfo{year}{1960}).

\bibitem[{\citenamefont{Laughlin}(1990)}]{Mclaug}
\bibinfo{author}{\bibfnamefont{J.~G.~M.} \bibnamefont{Laughlin}}, Ph.D. thesis,
  \bibinfo{school}{St. Hugh's College, Oxford} (\bibinfo{year}{1990}).

\bibitem[{\citenamefont{D\'ecanini and Folacci}(2008)}]{DecFol}
\bibinfo{author}{\bibfnamefont{Y.}~\bibnamefont{D\'ecanini}} \bibnamefont{and}
  \bibinfo{author}{\bibfnamefont{A.}~\bibnamefont{Folacci}},
  \bibinfo{journal}{Phys. Rev. D} \textbf{\bibinfo{volume}{78}},
  \bibinfo{pages}{044025} (\bibinfo{year}{2008}),
  \urlprefix\url{http://link.aps.org/doi/10.1103/PhysRevD.78.044025}.

\bibitem[{Mat(2007)}]{Mathematica}
\emph{\bibinfo{title}{Mathematica Edition}}, \bibinfo{address}{Champaign, IL,
  USA}, \bibinfo{edition}{version 7.0.1.0} ed. (\bibinfo{year}{2007}).

\bibitem[{\citenamefont{Ottewill and Wardell}(2009)}]{Barry:non-geo}
\bibinfo{author}{\bibfnamefont{A.~C.} \bibnamefont{Ottewill}} \bibnamefont{and}
  \bibinfo{author}{\bibfnamefont{B.}~\bibnamefont{Wardell}},
  \bibinfo{journal}{Phys. Rev. D} \textbf{\bibinfo{volume}{79}},
  \bibinfo{pages}{024031} (\bibinfo{year}{2009}).

\bibitem[{\citenamefont{Burkill}(1975)}]{Burkill}
\bibinfo{author}{\bibfnamefont{J.~C.} \bibnamefont{Burkill}},
  \emph{\bibinfo{title}{The Theory of Ordinary Differential Equations}}
  (\bibinfo{publisher}{Longman}, \bibinfo{year}{1975}).

\bibitem[{\citenamefont{Duffy and Ottewill}(2008)}]{DuffyOt}
\bibinfo{author}{\bibfnamefont{G.}~\bibnamefont{Duffy}} \bibnamefont{and}
  \bibinfo{author}{\bibfnamefont{A.~C.} \bibnamefont{Ottewill}},
  \bibinfo{journal}{Phys. Rev. D} \textbf{\bibinfo{volume}{77}},
  \bibinfo{pages}{024007} (\bibinfo{year}{2008}).

\bibitem[{\citenamefont{Breen}(2011)}]{PHD}
\bibinfo{author}{\bibfnamefont{C.}~\bibnamefont{Breen}}, Ph.D. thesis,
  \bibinfo{school}{University College Dublin, Dublin} (\bibinfo{year}{2011}).

\bibitem[{\citenamefont{Watson}(1944)}]{Watson}
\bibinfo{author}{\bibfnamefont{G.}~\bibnamefont{Watson}},
  \emph{\bibinfo{title}{A Treatise on the Theory of Bessel Functions}}
  (\bibinfo{publisher}{Cambridge University Press, Cambridge},
  \bibinfo{year}{1944}).

\bibitem[{\citenamefont{Gradshteyn and Ryzhik}(2000)}]{gradriz}
\bibinfo{author}{\bibfnamefont{I.}~\bibnamefont{Gradshteyn}} \bibnamefont{and}
  \bibinfo{author}{\bibfnamefont{I.}~\bibnamefont{Ryzhik}},
  \emph{\bibinfo{title}{Table of Integrals, Series and Products}}
  (\bibinfo{publisher}{Academic Press}, \bibinfo{year}{2000}).

\bibitem[{\citenamefont{Weniger}(2003)}]{Levrev}
\bibinfo{author}{\bibfnamefont{E.~J.} \bibnamefont{Weniger}},
  \bibinfo{journal}{ArXiv Mathematics e-prints}  (\bibinfo{year}{2003}),
  \eprint{arXiv:math/0306302}.

\bibitem[{\citenamefont{Balbinot et~al.}(2002)\citenamefont{Balbinot, Fabbri,
  Nicolini, and Sutton}}]{Babgen2d}
\bibinfo{author}{\bibfnamefont{R.}~\bibnamefont{Balbinot}},
  \bibinfo{author}{\bibfnamefont{A.}~\bibnamefont{Fabbri}},
  \bibinfo{author}{\bibfnamefont{P.}~\bibnamefont{Nicolini}}, \bibnamefont{and}
  \bibinfo{author}{\bibfnamefont{P.~J.} \bibnamefont{Sutton}},
  \bibinfo{journal}{Phys. Rev. D} \textbf{\bibinfo{volume}{66}},
  \bibinfo{pages}{024014} (\bibinfo{year}{2002}).

\bibitem[{\citenamefont{Morgan et~al.}(2007)\citenamefont{Morgan, Thom,
  Winstanley, and Young}}]{Winstanleygen}
\bibinfo{author}{\bibfnamefont{D.}~\bibnamefont{Morgan}},
  \bibinfo{author}{\bibfnamefont{S.}~\bibnamefont{Thom}},
  \bibinfo{author}{\bibfnamefont{E.}~\bibnamefont{Winstanley}},
  \bibnamefont{and} \bibinfo{author}{\bibfnamefont{P.}~\bibnamefont{Young}},
  \bibinfo{journal}{Gen. Rel. and Grav.}
  \textbf{\bibinfo{volume}{39}}, \bibinfo{pages}{1719} (\bibinfo{year}{2007}),
  ISSN \bibinfo{issn}{0001-7701}, \bibinfo{note}{10.1007/s10714-007-0486-3},
  \urlprefix\url{http://dx.doi.org/10.1007/s10714-007-0486-3}.

\bibitem[{\citenamefont{Gibbons and Hawking}(1977)}]{GibHawk}
\bibinfo{author}{\bibfnamefont{G.~W.} \bibnamefont{Gibbons}} \bibnamefont{and}
  \bibinfo{author}{\bibfnamefont{S.~W.} \bibnamefont{Hawking}},
  \bibinfo{journal}{Phys. Rev. D} \textbf{\bibinfo{volume}{15}},
  \bibinfo{pages}{2738} (\bibinfo{year}{1977}).

\bibitem[{\citenamefont{Riess et~al.}(1998)}]{Riess:1998}
\bibinfo{author}{\bibfnamefont{A.~G.} \bibnamefont{Riess}} \bibnamefont{et~al.}
  (\bibinfo{collaboration}{Supernova Search Team}), \bibinfo{journal}{Astron.
  J.} \textbf{\bibinfo{volume}{116}}, \bibinfo{pages}{1009}
  (\bibinfo{year}{1998}), \eprint{astro-ph/9805201}.

\bibitem[{\citenamefont{Matyjasek and Zwierzchowska}(2011)}]{Matyjasek:2011}
\bibinfo{author}{\bibfnamefont{J.}~\bibnamefont{Matyjasek}} \bibnamefont{and}
  \bibinfo{author}{\bibfnamefont{K.}~\bibnamefont{Zwierzchowska}}
  (\bibinfo{year}{2011}), \eprint{arXiv:1110.0041}.

\bibitem[{\citenamefont{Mellor and Moss}(1989)}]{Mellor}
\bibinfo{author}{\bibfnamefont{F.}~\bibnamefont{Mellor}} \bibnamefont{and}
  \bibinfo{author}{\bibfnamefont{I.}~\bibnamefont{Moss}},
  \bibinfo{journal}{Class. Quantum Grav.} \textbf{\bibinfo{volume}{6}},
  \bibinfo{pages}{1379} (\bibinfo{year}{1989}).

\bibitem[{\citenamefont{Murray}(1974)}]{Murray}
\bibinfo{author}{\bibfnamefont{J.~D.} \bibnamefont{Murray}},
  \emph{\bibinfo{title}{Asymptotic Analysis}} (\bibinfo{publisher}{Clarendon
  Press- Oxford}, \bibinfo{year}{1974}).

\bibitem[{\citenamefont{Dingle}(1973)}]{Dingle}
\bibinfo{author}{\bibfnamefont{R.}~\bibnamefont{Dingle}},
  \emph{\bibinfo{title}{Asymptotic Expansions: Their Derivation and
  Interpretation}} (\bibinfo{publisher}{Academic Press, London and New York},
  \bibinfo{year}{1973}).

\end{thebibliography}
\end{document}